\documentclass[11pt]{article}
\pdfoutput=1
\usepackage{jheppub}
\usepackage[T1]{fontenc}
\usepackage{diagbox}
\usepackage{adjustbox}
\usepackage{multirow}
\usepackage{pifont}
\usepackage{array}
\usepackage{mathtools}
\usepackage{booktabs}
\usepackage[dvipsnames]{xcolor}
\usepackage{tcolorbox}
\usepackage{float}
\usepackage{placeins}

\usepackage{verbatim}

\usepackage{graphics}
\usepackage{graphicx}
\usepackage{epsf}
\usepackage{epsfig}
\usepackage{float}
\usepackage{makecell}
\usepackage{multirow}
\usepackage{color}
\usepackage{xcolor}
\usepackage{simplewick}
\usepackage{longtable}

\usepackage[utf8]{inputenc}
\usepackage{amsmath}
\usepackage{amssymb}
\usepackage{subfig}
\usepackage[normalem]{ulem}

\usepackage{mathtools}
\usepackage{amsmath}
\usepackage{diagbox}

\usepackage{bm}       
\usepackage{bbm} 

\usepackage{relsize}  

\usepackage{autobreak}

\usepackage{makeidx} 
\usepackage{bbding} 
\usepackage{listings} 
\usepackage{ytableau}
\usepackage[utf8]{inputenc}
\usepackage[T1]{fontenc}
\usepackage{lmodern}

\usepackage{tikz}
\usepackage[vcentermath]{youngtab}
\usepackage{slashed} 
\usepackage[font=small]{caption} 
\usepackage{times}

\newcommand{\beq}{\begin {equation}}  
\newcommand{\eeq}{\end   {equation}} 
\newcommand{\bea}{\begin {eqnarray}} 
\newcommand{\eea}{\end   {eqnarray}}  
\newcommand{\baa}{\begin {array}   } 
\newcommand{\eaa}{\end   {array}   }     
\newcommand{\bit}{\begin {itemize} }
\newcommand{\eit}{\end   {itemize} }
\newcommand{\be }{\begin {equation}} 
\newcommand{\ee }{\end   {equation}}

\newcolumntype{M}[1]{>{\centering\arraybackslash}m{#1}}
\newcolumntype{N}{@{}m{0pt}@{}}

\allowdisplaybreaks

\newcommand{\lra}[1]{\langle #1 \rangle}

\newcommand{\bfu}{\mathbf{U}}
\newcommand{\bft}{\mathbf{T}}
\newcommand{\bfv}{\mathbf{V}}

\title{\boldmath Hilbert Series and Operator Counting on the Higgs Effective Field Theory}

\author[a,b]{Hao Sun,}
\author[a,b]{Yi-Ning Wang,}
\author[a,b,c,d,e]{Jiang-Hao Yu}

\affiliation[a]{CAS Key Laboratory of Theoretical Physics, Institute of Theoretical Physics, Chinese Academy of Sciences, Beijing 100190, China}
\affiliation[b]{School of Physical Sciences, University of Chinese Academy of Sciences, Beijing 100049, P.\ R.\ China}
\affiliation[c]{Center for High Energy Physics, Peking University, Beijing 100871, China}
\affiliation[d]{School of Fundamental Physics and Mathematical Sciences, Hangzhou Institute for Advanced
Study, UCAS, Hangzhou 310024, China}
\affiliation[e]{International Centre for Theoretical Physics Asia-Pacific, Beijing/Hangzhou, China}

\emailAdd{sunhao@itp.ac.cn}
\emailAdd{wangyining@itp.ac.cn}
\emailAdd{jhyu@itp.ac.cn}

\abstract{We present a systematic procedure for determining the Hilbert series that counts the number of independent operators in the Higgs effective field theory. After removing the redundancies from equation-of-motion and integration-by-part, we provide an algorithm of treating the redundancy from the operators involving in spurion fields parametrizing the custodial symmetry breaking. Furthermore, we utilize the outer automorphism of the Lorentz and internal symmetries to separate operators with different CP properties. With these new implements, the Hilbert series up to chiral dimension 10 are obtained, and CP-even and CP-odd operators can be further divided. Extensions to all orders are straightforward. 
}

\begin{document} 
\maketitle
\flushbottom

%
%
%

\newpage
\section{Introduction}

The Standard Model (SM) of the particle physics has been acknowledged as the most successful theory that explains a wide variety of phenomena with unprecedented accuracy, yet several experimental facts motivate new physics beyond the SM. 
The failure of discovering these new particles at the Large Hadron Collider (LHC) below few TeVs suggests that there is a considerable energy gap between the SM particles and new physics. Under this circumstance, the effective field theory (EFT) provides a new paradigm to investigate the new physics effects systematically, instead of traditional model building of new physics. 
Pioneered by Weinberg~\cite{Weinberg:1978kz}, EFT has been developed to be a systematic framework in many areas of particle physics, among which the most fruitful one is the standard model effective field theory (SMEFT). Since the dimension-5
operators were first written down by Weinberg~\cite{Weinberg:1979sa}, many progresses have been made~\cite{Weinberg:1979sa,Grzadkowski:2010es,Lehman:2014jma,Henning:2015alf,Liao:2016hru,Li:2020xlh,Murphy:2020rsh,Li:2020gnx,Liao:2020jmn,Li:2022tec}. Recently a general algorithm, implemented in a Mathematica package ABC4EFT~\cite{Li:2022tec}, was proposed to construct the independent and complete SMEFT operators up to any mass dimension.

Above the electroweak scale, the SMEFT parametrizes all possible Lorentz-invariant new physics satisfying the SM gauge symmetry. Except the baryon and lepton number, there is another accidental global symmetry in the Higgs sector: the custodial symmetry $\mathcal{G}=SU(2)_L\times SU(2)_R$. Below the electroweak scale, the custodial symmetry is spontaneously broken to the subgroup $\mathcal{H}=SU(2)_V$ via the vacuum expectation value (vev) of the Higgs fields. The spontaneous symmetry breaking $\mathcal{G} \to \mathcal{H}$ coset pattern can be described by the Callan-Coleman-Wess-Zumino (CCWZ) formalism~\cite{Callan:1969sn,Coleman:1969sm}, similar to the chiral perturbation theory (ChPT)~\cite{Gasser:1983yg,Gasser:1984gg} with the chiral dimension $d_\chi$ power counting. The electroweak chiral Lagrangian (EWChL) has been investigated in Ref.~\cite{Longhitano:1980iz,Longhitano:1980tm,Appelquist:1980vg,Dobado:1989ax,Feruglio:1992wf,Herrero:1993nc,Herrero:1994iu}. After the Higgs discovery, the electroweak chiral Lagrangian with the light Higgs is revived as the Higgs effective field theory (HEFT), and the operators up to next-to-leading order (NLO) has been investigated in literature~\cite{Buchalla:2012qq,Alonso:2012px,Buchalla:2013rka,Brivio:2013pma,Gavela:2014vra,Pich:2015kwa,Brivio:2016fzo,Pich:2016lew,Merlo:2016prs,Krause:2018cwe,Pich:2018ltt}. Recently the complete and independent NLO operators of the HEFT has been written in Ref.~\cite{Sun:2022ssa}, in which 6(9) new terms of operators were identified and many redundancies among operators were removed based on the Young tensor technique developed in Ref.~\cite{Li:2020xlh,Li:2020gnx,Li:2022tec}.


Compared to the SMEFT, the HEFT Lagrangian consists of the Nambu-Goldstone bosons (NGBs) from the $\mathcal{G}/ \mathcal{H}$ coset, and the physical Higgs boson as the electroweak singlet. Thus the HEFT provides more general parametrization on the Higgs sector than the SMEFT. This fact can be seen when matching the SMEFT operators into the HEFT ones. The dimension 6 and 8 SMEFT operators should be matched to the NLO and next-to-next-to-leading order (NNLO) operators of the HEFT. Recently the NNLO Lagrangian was constructed for the first time in Ref.~\cite{Sun:2022snw}, with the completeness and independence guaranteed.

Apart from the explicit form of the effective operators, it is also important to enumerate the numbers of the independent operators at each order, which can be done by the Hilbert series method. The Hilbert series is an algebra technique to counting the group invariants, which utilises the group characters and the Schur's lemma. Although it can not offer the explicit form of  operators, it provides an independent cross-check on the operator basis. 
The Hilbert series method was first introduced to enumerate independent gauge invariants~\cite{Feng:2007ur,Jenkins:2009dy,Hanany:2010vu,Hanany:2014dia}. Then the Hilbert series is utilized in the SMEFT~\cite{Lehman:2015via, Lehman:2015coa,Henning:2015daa,Henning:2015alf,Henning:2017fpj,Marinissen:2020jmb} to count the independent operators by removing the redundancies from equation of motion (EOM) and integration-by-part (IBP). Now this method has been extended to many other EFTs, such as the flavor invariant~\cite{Jenkins:2009dy,Hanany:2010vu,Wang:2021wdq,Yu:2021cco,Bonnefoy:2021tbt,Yu:2022ttm}, non-relativistic QCD~\cite{Kobach:2017xkw}, beyond the SM model~\cite{Anisha:2019nzx}, the gravity EFT~\cite{Ruhdorfer:2019qmk}, QCD chiral Lagrangian~\cite{Graf:2020yxt}, and so on.

In this work, we apply the Hilbert series technique to count the number of invariants in the HEFT, partly to cross-check the NLO and NNLO results in Ref.~\cite{Sun:2022ssa, Sun:2022snw}~\footnote{We find the agreement on the numbers of the HEFT operator basis between the Young tensor technique~\cite{Sun:2022ssa, Sun:2022snw} and the Hilbert series in this work. }, and partly to extend the NLO and NNLO results to all orders. We adapt the same treatments on the equation-of-motion and integration-by-part redundancies as Ref.~\cite{Lehman:2015coa,Henning:2015alf}. Furthermore, there are additional new treatments:
\bit
\item In the HEFT, the spurion $\bft$ parametrizing the custodial symmetry breaking effects, is needed to form group invariants. The spurion behaves as frozen degree of freedom, and arises new redundancy of the effective operators. We extend the Hilbert series method to manage this special building blocks via its factorization property. 
\item
Besides, the CP properties of the HEFT operators is also quite important. We investigate the outer automorphism of the Lorentz and internal symmetries to extend the Hilbert series to four different CP branches. Thus it is possible to count the numbers of the operators with different CP structures in the HEFT. 
\eit
After these extensions, the Hilbert series can give the numbers of independent operators and their CP structures up to all orders. 

In this paper, we briefly introduce the Hilbert method in the Sec.~\ref{sec:HS}, including the elimination of the redundancy from spurions and the CP structures. Next, we review the HEFT in Sec.~\ref{sec:HEFT}, and present the numbers of independent operators up to chiral dimension 10 and identify the CP structures. At last we conclude this paper in Sec.~\ref{sec:conc}.

\section{Hilbert Series}
\label{sec:HS}

\subsection{Plethystic Exponential}

Hilbert series $\mathcal{H}(q)$ is a generating function that counts the number of independent group invariants that can be formed from a set of multiplets $q_i$ in different representations ${\bf R}_i$ of the group $G$ via a power series
\bea
\mathcal{H}(q_i) = \sum_{k_1 = 0,\cdots,k_n=0}^{\infty} c_{k_1,...,k_n} q_{1}^{k_{1}}...q_{n}^{k_{n}}\,,
\eea
where $n$ denotes the species of the multiplets, and $c_{k_1,...,k_n}$ is the number of invariants of order $k_i$ in $q_i$ with $i = 1, \cdots, n$. According to group theory, $c_{k_1,...,k_n}$ denotes the multiplicity of the trivial representation of the tensor product ${\text{sym}^{k_{1}}\bf R}_{1}\otimes \cdots \otimes {\text{sym}^{k_{n}}\bf R}_{n}$ under the group $G$, and can be projected out using the ortho-normal property of the group character $\chi_{\bf R_i}(g), g \in G$.

In the field theory setup, if there are $N$-field operators in the Lorentz-invariant theory, the multiplets are taken to be all the fields $\phi_i$ and the (covariant) derivatives $D$, furnishing certain representations $\mathbf{R}$ of the group $G$, which includes Lorentz group and possibly internal symmetry groups. 
The invariant Lagrangian can be described by the Hilbert series, which takes the form
\begin{equation}
    {\mathcal H}(D,{\{\phi_{i}\}})=\sum_{k_{0}=0}^{\infty}\sum_{k_{1}=0}^{\infty}...\sum_{k_{N}=0}^{\infty} c_{k_0,k_1,...,k_N}\phi_{1}^{k_{1}}...\phi_{N}^{k_{N}}D^{k_{0}}
    \,,
\end{equation}
where $\phi_{i}$ is only a variable instead of the field operator itself. 
According to the spin-statistics of the field operators, the bosonic (fermionic) field should satisfy the (anti-)commutation relation, thus only the (anti-)symmetric tensor product $\text{sym}^n(\phi)$ ($\text{anti-sym}^n(\phi)$) should be involved in the Hilbert series. 
%
%

To identify the $q_i$ in the field theory settings, the single particle module (SPM)\cite{Henning:2017fpj} of $\phi$ is defined
\bea
R_{\phi} = \left(
\begin{array}{c}
\phi\\
D_{\mu}\phi\\
D_{(\mu_{1}}D_{\mu_{2})}\phi \\
\vdots
\end{array}
\right).
\eea
Here $D_{(\mu_{1}}D_{\mu_{2})}$ is the symmetric part of $D_{\mu_{1}}D_{\mu_{2}}$. The anti-symmetric parts disappear because they 
can be converted into other operators via the covariant derivative commutators and do not contribute to the number of independent physical operators. So we identity $D_{\mu_{1}}D_{\mu_{2}}$ with its symmetric part $D_{(\mu_{1}}D_{\mu_{2})}$ in this paper. The character $\chi_{_{R_{\phi}}}(D,g)$ of the $R_{\phi}$, which is the function of $D$ and group element $g\in G$, is the sum of the character of the field with $n$-derivative $\text{sym}^n(D)\phi$:
\begin{equation}
\label{kfactor}
    \chi_{_{R_{\phi}}}(D,g)=\sum_{n}^{\infty}D^n \chi_{\text{sym}^n(D)}(g)\chi_{\bf R}(g)\,,\quad g\in G
    \,.
\end{equation} 
The character of $n$ repeated modulus are summed by the plethystic exponential (PE)~\cite{Feng:2007ur}, which takes the form 
\bea
    \text{PE}[\phi\chi_{_{R_{\phi}}}(D,g)] &\equiv& \exp \left[\operatornamewithlimits{\sum}_{n=1}^{\infty}\frac{\phi^{n}}{n}\chi_{_{R_{\phi}}}(D^n,g^{n})\right]=\,\operatornamewithlimits{\sum}_{n=1}^{\infty}\phi^{n}\chi_{\text{sym}^{n}(R_{\phi})}(D,g)  \\ 
    \text{PE}_{f}[\phi\chi_{_{R_{\phi}}}(D,g)]&\equiv& \exp \left[ - \operatornamewithlimits{\sum}_{n=1}^{\infty}\frac{(-\phi)^{n}}{n}\chi_{_{R_{\phi}}}(D^n,g^{n})\right]=\, \operatornamewithlimits{\sum}_{n=1}^{\infty}\phi^{n}\chi_{\text{anti-sym}^{n}(R_{\phi})}(D,g)\,,
    \label{eq:PE}
\eea
for bosons and fermions respectively. Using the Schur's Lemma, the Hilbert series is obtained by projecting the PE onto the trivial representation
\begin{equation}
    {\mathcal{H}}(D,\phi)=\int_{G} d\mu(g) \left\{   \begin{array}{ll}
    \text{PE}[\phi\chi_{_{R_{\phi}}}(D,g)] & \phi \text{ is boson}   \\
    \text{PE}_{f}[\phi\chi_{_{R_{\phi}}}(D,g)]  &\phi \text{ is fermion}\,.
    \end{array}
    \right.
    \label{eq:HSphi}
\end{equation}
Here $d\mu$ is the Haar measure for the group $G$. The Haar measure of some groups which will be used in this work are listed in Table.~\ref{tab:char}.


For $N$ multiple fields $\{\phi_{i}\}$ one multiplies their PEs altogether
\begin{equation}
    {\mathcal H}(D,\{\phi_{i}\})=\int_{G} d\mu(g) \operatornamewithlimits{\Pi}_{i=1}^{N}\left\{
    \begin{array}{ll}
    \text{PE}[\phi_{i}\chi_{_{R_{\phi_{i}}}}(D,g)] & \phi_{i} \text{ is boson}   \\
    \text{PE}_{f}[\phi_{i}\chi_{_{R_{\phi_{i}}}}(D,g)]  &\phi_{i} \text{ is fermion}\,.
    \end{array}
    \right.
    \label{eq:HSphi2}
\end{equation}
After performing integration over group elements and rescaling the fields according to their mass dimension
\bea
\phi_i \to \epsilon^d \phi_i, \quad D \to \epsilon D, 
\eea
the final form of the Hilbert series is a formal series of the variable $\phi_i$, organized in terms of a series in mass dimension, corresponding to the effective operators organized by the same mass dimension. 

The Hilbert series above corresponds to the effective operators containing the equation-of-motion (EOM) and the integration-by-part (IBP) redundancies. 
Let us remove these redundancies by modifying the module characters and extracting out the primary operators. 
%
%

\subsection{EOM Redundancy}
\label{sec:IBP}

Inside the single particle module $R_{\phi}$, the field with $n$-derivative $\text{sym}^n(D)\phi$ can be decomposed into irreducible representations (irreps) of the Lorentz group. For a scalar with $n$-derivative $\text{sym}^n(D) \varphi$ in $R_{\varphi}$, the decomposition takes
\begin{equation}
    \text{sym}^{n}(\frac{1}{2},\frac{1}{2})=(\frac{n}{2},\frac{n}{2})\oplus(\frac{n}{2}-1,\frac{n}{2}-1)\oplus...
    \,,
\end{equation}
where $(\frac{n}{2},\frac{n}{2})$ is the symmetric and traceless part of $\text{sym}^n(D)$. The other representations $(\frac{n}{2}-k,\frac{n}{2}-k)$ ($k>0$) should be removed because they always contains $D^2\varphi$ from the trace contraction in the derivatives, and thus the EOM $(D^2+m^2)\varphi=0$ in the module is removed. For example, 
\begin{equation}
    D_{\mu}D_{\nu}\varphi=(D_{\mu}D_{\nu}-\frac{1}{4}D^2\eta_{\mu\nu})\varphi+\frac{1}{4}\eta_{\mu\nu}D^2\varphi
    \,,
\end{equation}
we note that the second term on the right side is the EOM. 
Keeping only the symmetric and traceless part, such as $D_{\{\mu_{1}}D_{\mu_{2}\}}$, the scalar module is reduced to 
\bea
\label{eq:rphi}
R_{\varphi} = \left(
\begin{array}{c}
\varphi\\
D_{\mu}\varphi\\
D_{\{\mu_{1}}D_{\mu_{2}\}}\varphi \\
\vdots
\end{array}
\right)
\,.
\eea
The character of the $R_{\varphi}$ becomes
\bea
        \chi_{_{R_{\varphi}}}(D,g) &= \sum_{n}^{\infty}D^n \chi_{_{(\frac{n}{2},\frac{n}{2})}}(g)\chi_{_{(0,0)}}(g)  
             = (1-D^2)P(D,g)\chi_{_{(0,0)}}(g)\,,
\eea
where $P(D,g)$ denotes the character of the symmetrical derivatives' tensor product. The Lorentz group character of certain element $g$ in certain representation $\chi_{(j_1,j_2)}(g)$ can be parameterized by its maximal torus variables $a=(a_1,a_2)$ as $\chi_{(j_1,j_2)}(a)$\footnote{$a$ are actually the maximal torus variables of $SO(4,\mathbb{C})$, since the characters are the functions of equivalent classes and the Lie algebras of the two groups are isomorphic.}, and the explicit expressions of some irreps can be found in Table.~\ref{tab:char}. As a consequence, the factor $P$ can be expressed by $a$ as
\bea
P(D,g(a)) \equiv \frac{1}{(1-Da_1)(1-D/a_1)(1-Da_2)(1-D/a_2)} \,.
\label{eq:kinematical factor}
\eea


For the left-handed fermion module, the decomposition on a general $\text{sym}^n(D) \psi_L$ takes the form
\begin{equation}
\label{fermEOM}
    \text{sym}^{n}(\frac{1}{2},\frac{1}{2}) \otimes (\frac{1}{2},0) =(\frac{n+1}{2},\frac{n}{2})\oplus \cdots\,.
\end{equation}
Here similarly the highest part $(\frac{n+1}{2},\frac{n}{2})$ is kept, and the part other than it contains the $D\!\!\!\!/\psi_{L}\sim(0,\frac{1}{2})$ terms and should be removed.  
%
After removing all the other parts in the module, the character of the module $R_{\psi_{L}}$ becomes
\begin{equation}
    \chi_{_{R_{\psi_{L}}}}(D,a)=\left(\chi_{_{(\frac{1}{2},0)}}(a)-D\chi_{_{(0,\frac{1}{2})}}(a)\right)P(D,a).
\end{equation}
For the left-handed field strength $F_{L}$ with $n$-derivatives, the decomposition takes
\begin{equation}
    \text{sym}^{n}(\frac{1}{2},\frac{1}{2})\otimes(1,0)=(\frac{n+2}{2},\frac{1}{2})\oplus \cdots
\end{equation}
By reserving the highest part $(\frac{n+2}{2},\frac{1}{2})$, the EOM $D_{\mu}F_{L}^{\mu\nu}=J^{\nu}$ redundancy is removed, equivalently the contribution of $J^{\nu}$ is subtracted in the field strength SPM. The current conservation $D_{\mu}J^{\mu}=0$ indicates that the character of the SPM for $D_{\mu}F^{\mu\nu}$ takes
\begin{equation}
    \chi_{_{R_{_{D_{\mu}F^{\mu\nu}}}}}(D,a)=D\left(\chi_{_{(\frac{1}{2},\frac{1}{2})}}(D,a)-D\chi_{_{(0,0)}}(D,a)\right)P(D,a).
\end{equation}
Accordingly the character of $F_{L}$ SPM after all kinds of subtractions is
\begin{equation}
        \chi_{_{R_{F_{L}}}}(D,a)=\left(\chi_{_{(1,0)}}(D,a)-D\chi_{_{(\frac{1}{2},\frac{1}{2})}}(D,a)+D^2\chi_{_{(0,0)}}(D,a)\right)P(D,a).
\end{equation}
%
Similarly for the right-handed fermions and vector bosons.

\subsection{IBP Redundancy}

The IBP redundancy identifies the operators related by total-derivative terms.
After removing the EOM for each SPM, the modules $\{ R_{\phi_{1}}, \cdots, R_{\phi_{N}}\}$ would form a set of effective operators.  
If the constructed operator set can be organized in the form similar to the SPM
\begin{equation}
R_{\mathcal{O}_i}=\left(
\begin{array}{c}
     \mathcal{O}_i  \\
     D\mathcal{O}_i  \\
     D^2\mathcal{O}_i \\
     \vdots
\end{array}
\right)\,,
\end{equation}
the IBP redundancy corresponds to all the descendants of $\mathcal{O}_i$ in $R_{\mathcal{O}_i}$. Conversely, by removing all the descendants the IBP redundancy immediately gets removed. 
This procedure was firstly proposed in \cite{Henning:2017fpj} by realizing that the module set $\{ R_{\phi_{1}}, \cdots, R_{\phi_{N}}\}$~\footnote{According to group theory, removing EOM in SPMs corresponds to dropping the negative-norm states in the unitary representation of the conformal group~\cite{Minwalla1997RestrictionsIB,Ferrara:2000nu,Dolan:2005wy}. } actually correspond to the unitary representation space of the conformal group $SO(4,2,\mathbb{C})\simeq SO(6,\mathbb{C})$. 
Thus a set of operators formed by SPMs would furnish the tensor-product representations of the conformal group, which can be decomposed into irreps in the $\{ R_{\mathcal{O}_i}\}$ space
\bea
\oplus_{n_1, \cdots n_N} \left[ {\textrm{sym}}^{n_1}\left(R_{\phi_{1}}\right)\otimes \cdots \otimes {\textrm{sym}}^{n_N}\left(R_{\phi_{N}}\right)  \right]=\oplus_i R_{\mathcal{O}_{i}}\,.
\eea
In each irrep $\{ R_{\mathcal{O}_i}\}$, the $\{\mathcal{O}_i\}$ part corresponds to the primary states\footnote{Here primary means that the hermitian conjugation $K_{i}$ of momentum $P_{i}$ act on those states gives zero hence they are not images of $P_{i}$.} of the unitary conformal representation, while other descendant parts corresponding the descendant states. Once the decomposition is done, the $\{\mathcal{O}_i\}$ part in $\{ R_{\mathcal{O}_i}\}$ is obtained as the operator set with the IBP removed. In $SO(6,\mathbb{C})$ the irrep $\{ R_{\mathcal{O}_i}\}$ is labelled by $[\Delta_0,j_{1},j_{2}]$ with scaling dimension $\Delta_0$ and spin $(j_{1},j_{2})$ of $\{\mathcal{O}_i\}$. The corresponding maximal torus variables are denoted by $q$ and $a=(a_{1},a_{2})$.
%
%
%

Since the module $R_{\mathcal{O}}$ corresponds to an irreps in $SO(6,\mathbb{C})$ labelled by $[\Delta_{0},j_{1},j_{2}]$, the character $\mathcal{X}_{[\Delta_0,j_{1},j_{2}]}$ in $SO(6,\mathbb{C})$ can be directly obtained from the character $\chi_{R_{\mathcal{O}}}(D,a)$ in $SO(4,\mathbb{C})$ by changing $D$ into the scaling factor $q$ and multiplying the scaling factor $q^{\Delta_{0}}$ of $\mathcal{O}$ on the character in $SO(4,\mathbb{C})$
\begin{equation}
    {\mathcal{X}}_{[\Delta_0,j_{1},j_{2}]}(q,a)=q^{\Delta_0}\chi_{_{R_{\mathcal{O}}}}(q,a)\,,
\end{equation}
where $\Delta_{0}$ is the scaling dimension of $\mathcal{O}$. 
The character satisfies the ortho-normal relation using its dual character $\chi_{[\Delta_0,j_{1},j_{2}]}^{\text{ortho}}$
\begin{equation}
    \int\frac{dq}{2\pi iq}\frac{d\mu_{L}(a)}{|P(q,a)|^2}\mathcal{X}_{[\Delta_0',j'_{1},j'_{2}]}^{\text{ortho}*}\mathcal{X}_{[\Delta_0,j_{1},j_{2}]}=\delta_{\Delta_0',\Delta_0}\delta_{j'_{1},j_{1}}\delta_{j'_{2},j_{2}}\,.
    \label{orth}
\end{equation}
Here $d\mu_{L}(a)$ and $\frac{dq\, d\mu_{L}(a)}{2\pi iq|P^+(q,a)|^2}$ are the Haar measure over the maximal torus of the $SO(4,\mathbb{C})$ and the $SO(6,\mathbb{C})$ group, respectively.

Under the conformal group $SO(6,\mathbb{C})$, the Hilbert series $\mathcal{H}(D,\phi)$ in Eq.~\ref{eq:HSphi} should be modified because the Lorentz scalars are trivial representations in $SO(4,\mathbb{C})$ but correspond to non-trivial representations in $SO(6,\mathbb{C})$. So the Hilbert series should be obtained by projecting the PE onto irreps $[\Delta_{0},0,0]$ instead of the trivial one. Using the relation Eq.~\ref{orth}, the Hilbert series is re-expressed as  
\bea
\mathcal{H}(D,\phi)&=& 1+\operatornamewithlimits{\sum}_{\Delta_0}D^{\Delta_0} \int\frac{dq}{2\pi iq}\frac{d\mu_{L}}{|P(q,a)|^2}\mathcal{X}_{[\Delta_0,0,0]}^{\text{ortho}*}\left[
    \text{PE}(\phi q^{\Delta_{\phi}}D^{-\Delta_{\phi}} \chi_{_{R_{\phi}}}(q,a))-1
    \right]\\
    &=& \int d\mu_{L}(a)\frac{1}{P(D,a)}\text{PE}\left(
    \phi \chi_{_{R_{\phi}}}(D,a)
    \right) +\Delta \mathcal{H}\,.
\eea
Here $\Delta_{\phi}$ is the mass dimension of $\phi$, and a rescaling $\phi\rightarrow\phi D^{-\Delta_{\phi}}$ has been performed to keep the power of $D$ still counting the number of derivatives. $\Delta \mathcal{H}$ in the second line only contributes to the relevant terms ($\Delta_0 \leq 4$)\cite{Henning:2017fpj} hence we do not concentrate on it and just drop it out in the final Hilbert series.

To include the multiple fields cases, we multiply the corresponding PE($\text{PE}_{f}$) functions in the integral\cite{Lehman:2015via}. And to incorporate the internal symmetry, we need to perform integration over internal group and  replace the character $\chi_{_{R_{\phi}}}(q,a)$ of $SO(4)$ by
\bea
\chi_{_{R_{\phi}}}(q,a,b)=\chi_{_{R_{\phi}}}(q,a)\chi_{_{\text{Internal}}}(b)
\label{eq:character}
\eea
where $b$ parameterizes the internal symmetry. The final form is obtained 
\begin{equation}
    {\mathcal H}(D,\{\phi_{i}\})=\int_{G} d\mu(a,b)\frac{1}{P(D,a)} \operatornamewithlimits{\Pi}_{i=1}^{N}\left\{
    \begin{array}{ll}
    \text{PE}[\phi_{i}\chi_{_{R_{\phi_{i}}}}(D,a,b)] & \phi_{i} \text{ is boson}   \\
    \text{PE}_{f}[\phi_{i}\chi_{_{R_{\phi_{i}}}}(D,a,b)]  &\phi_{i} \text{ is fermion}\,.
    \end{array}
    \right.
    \label{eq:finalHS}
\end{equation}
In summary, the IBP redundancy is removed by dividing a simple factor $P(D,a)$ in Eq.~\ref{eq:HSphi2} and concentrating on the irrelevant operators ($\Delta_0 >4$).

\subsection{Spurion Field}
\label{sec:spurion}
In some field theory, such as Higgs EFT, some fields get vacuum expectation value (vev) at low energy and behave like frozen degrees of freedom, thus the operators with the derivatives applying on such fields vanish. These fields are so-called spurions, which carry no dynamic information but are needed to form invariants under the internal group. Due to such special property, its SPM is extremely truncated
\begin{equation}
    R_{\mathbf{T}} = \left(\bft\right)\,,
\end{equation}
since the derivatives applying on it gives zero, where we have used $\mathbf{T}$ representing spurion. According to the discussion in Sec.~\ref{sec:IBP}, the character of such SPM takes the form that
\begin{equation}
    \chi_{_{R_\mathbf{T}}}(D,a,b) = \underbrace{\chi_{_{(0,0)}}(a)}_{=1}\chi_{_{\text{Internal}}}(b)\,,
\end{equation}
by which, the operators with derivatives applying on the spurions are eliminated.

For operators containing spurions $\bft$ and $n$ fields $\phi$, there are additional redundant relations among different types of operators, such that 
\bea
\label{eq:ren}
\left[\bft^{\otimes k}\right] \bft^m \phi^n 
\sim \bft^m \phi^n
\eea
where $\bft^m \phi^n$ represents any effective operator containing the spurions, and $\left[\bft^{\otimes k}\right]$ denotes the group invariant constructed by the tensor product of $k$ spurions $\bft$. This group invariants composed purely by $\bft$ can also be counted via the Hilbert series as
\begin{equation}
    \mathcal{H}(\bft)=\sum_{k=0}^\infty f_k\bft^k\,,
\end{equation}
where $f_k$ is the number of the invariants $\left[\bft^{\otimes k}\right]$ and $f_0=1$.

%

The complete Hilbert series of some fields $\phi_i$ and spurion $\bft$ can be organised in terms of number of spurions,
\begin{equation}
    \mathcal{H}^{\text{sp}}(D,\{\phi_i\},\bft) = \sum_{m=0}^\infty \mathcal{H}^{\text{sp}}(D,\{\phi_i\},\bft^{\otimes m})\,,
\end{equation}
where the superscript "sp" implies it contains the redundancy in Eq.~\ref{eq:ren}, and $\bft^{\otimes m}$ denotes the tensor product of $m$ spurions. 
%
If there are $k$ spurions in the type of operators, the spurions in $\mathcal{H}^{\text{sp}}(D,\{\phi_i\},\bft^{\otimes k})$ do not necessarily form a singlet, but there should have operators constituted by the following possible singlets
\bea
\{ \left[\bft^{\otimes 1}\right], \left[\bft^{\otimes 2}\right], \cdots,  \left[\bft^{\otimes k}\right]\}
\eea
The operators containing the above pieces should be eliminated due to the spurion redundancy relation in Eq.~\ref{eq:ren}.  
%

In the following we present an algorithm to eliminate this redundancy. Denote the physical Hilbert series as
\begin{equation}
    \mathcal{H}(D,\{\phi_i\},\bft) = \sum_{m=0}^\infty \mathcal{H}(D,\{\phi_i\},\bft^{\otimes m})\,,
\end{equation}
where the first term in the right side is the same with the first one of $\mathcal{H}^{\text{sp}}(D,\{\phi_i\},\bft) $, such that
\begin{equation}
\label{eq:k=0}
     \mathcal{H}^{\text{sp}}(D,\{\phi_i\},\bft^{\otimes 0}) = \mathcal{H}(D,\{\phi_i\},\bft^{\otimes 0})\,.
\end{equation}
Considering the general term of $k$ spurions,
the complete Hilbert series $\mathcal{H}^{\text{sp}}(D,\{\phi_i\},\bft^{\otimes k})$ can be decomposed into several disjoint parts: the physical part 
$$\mathcal{H}(D,\{\phi_i\},\bft^{\otimes k}),$$ 
the product of the physical part of $(k-1)$-spurions case and the 1-spurions term in $\mathcal{H}(\bft)$, 
$$\mathcal{H}(D,\{\phi_i\},\bft^{\otimes k-1})\times f_1\bft\,,$$
the product of the physical part of $(k-2)$-spurions case and the 2-spurions term in $\mathcal{H}(\bft)$, 
$$\mathcal{H}(D,\{\phi_i\},\bft^{\otimes k-2})\times f_2\bft^2$$
and so on, thus we can prove the general relation among these 3 different Hilbert series by induction of the spurion number $k$,
\begin{equation}
\label{eq:m=k}
\mathcal{H}^{\text{sp}}(D,\{\phi_i\},\bft^{\otimes k}) = \sum_{m=0}^k\mathcal{H}(D,\{\phi_i\},\bft^{\otimes m})f_{k-m}\bft^{k-m}\,,
\end{equation}
and the special case of $k=0$ is just Eq.~\ref{eq:k=0}. Equivalently, the physical Hilbert series with $k$ spurions can be obtained by the subtraction
\begin{equation}
    \mathcal{H}(D,\{\phi_i\},\bft^{\otimes k}) = \mathcal{H}^{\text{sp}}(D,\{\phi_i\},\bft^{\otimes k}) - \sum_{j=1}^{k}\mathcal{H}(D,\{\phi_i\},\bft^{\otimes k-j}) f_j\bft^j \,,\label{eq:subs}
\end{equation}
Combine the relations in Eq.~\ref{eq:m=k} of any $k$ together, we can get the complete result that
\begin{equation}
    \mathcal{H}^{\text{sp}}(D,\{\phi_i\},\bft) = \sum_{k=0}^\infty \sum_{m=0}^k\mathcal{H}(D,\{\phi_I\},\bft^{\otimes m})f_{k-m}\bft^{k-m} = \mathcal{H}(D,\{\phi_i\},\bft)\mathcal{H}(\bft)\,,
\end{equation}
or
\begin{equation}
    \mathcal{H}(D,\{\phi_i\},\bft)=\frac{ \mathcal{H}^{\text{sp}}(D,\{\phi_i\},\bft) }{\mathcal{H}(\bft)}\,.
\end{equation}
Thus we arrive the conclusion that the complete Hilbert series $\mathcal{H}^{\text{sp}}(D,\{\phi_i\},\bft)$ modulo the Hilbert series of spurion $\mathcal{H}(\bft)$ eliminates the new redundancy in Eq.~\ref{eq:ren}, which, when the infinite series is truncated to certain dimension, is just the subtraction as shown in Eq.~\ref{eq:subs}.

\subsection{Charge Conjugate and Parity}
\label{sec:cp}

 In a Lorentz-invariant field theory, the symmetry group $G=SO(4)\times I$ with $I$ representing the internal symmetry group determine the symmetry of the Lagrangian in most cases. In addition, there exist three extra discrete transformations: charge conjugation $\mathcal{C}$~\footnote{Here we use a different notation $\mathcal{C}$ with that $\mathcal{C}$ in the Ref.~\cite{Graf:2020yxt}, since $\mathcal{C}$ here changes the chiral property of particles but $\mathcal{C}$ in the reference does not. }, space inversion $\mathcal{P}$ and time reversal $\mathcal{T}$. Although they does not need to be symmetries of the theory, it is sometimes necessary to study them, such as the CP violation, etc. Due to the general CPT theorem in the field theory, only two of them are necessarily considered so we only consider $\mathcal{C}$ and $\mathcal{P}$ here. The actions of $\mathcal{C}$ and $\mathcal{P}$ on irreps $(j_{1},j_{2},\bf r)$ of the $G$ is listed in Figure.~\ref{CPaction}.
\begin{figure}[ht]
    \centering
    \begin{tikzpicture}
\node (1) at(0,0) {$(j_{1},j_{2},\bf r)$};
\node (2) at(4,0) {$(j_{1},j_{2},\bf r^{*})$};
\node (3) at(0,2) {$(j_{2},j_{1},\bf r)$};
\node (4) at(4,2) {$(j_{2},j_{1},\bf r^{*})$};
\node (5) at(2,0) {$\mathcal{CP}$};
\node (6) at(2,2) {$\mathcal{CP}$};
\node (7) at(0,1) {$\mathcal{P}$};
\node (8) at(4,1) {$\mathcal{P}$};
\node (9) at(2,1) {$\mathcal{C}$};
\draw[<-] (1) --(5);
\draw[->] (5) --(2);
\draw[<-] (1) --(5);
\draw[->] (5) --(2);
\draw[<-] (1) --(7);
\draw[->] (7) --(3);
\draw[<-] (1) --(7);
\draw[->] (7) --(3);
\draw[<-] (2) --(8);
\draw[->] (8) --(4);
\draw[<-] (2) --(8);
\draw[->] (8) --(4);
\draw[<-] (3) --(6);
\draw[->] (6) --(4);
\draw[<-] (3) --(6);
\draw[->] (6) --(4);
\draw[->] (9) --(1);
\draw[->] (9) --(2);
\draw[->] (9) --(3);
\draw[->] (9) --(4);
\end{tikzpicture}
    \caption{The actions of $\mathcal{C}$ and $\mathcal{P}$ on representations of $G$. Here $\bf r$ is irrep of $I$ and $\bf r^{*}$ is its complex conjugation representation. $(j_{1},j_{2})$ label the representations of $SO(4)$.}
    \label{CPaction}
\end{figure}

To consider the CP properties of effective operators, one can add $\mathcal{C}$ and $\mathcal{P}$ into the symmetry group $G$ by thinking them as the outer automorphism of $G$\cite{Grimus:1995zi,Buchbinder:2000cq}. For the Lorentz group, it becomes $SO(4)\rtimes \{1,\mathcal{P}\}=O(4)$. This is the disjoint union of two connected components $O(4)=SO(4)\sqcup O_{-}(4)$ with $O_{-}(4)$ the parity-odd component. Similarly, for the internal group it becomes $\tilde{I}\equiv I\rtimes{\{1,\mathcal{C}\}}=I\sqcup I_{-}$.
For now the total group is $\tilde{G}=O(4)\rtimes\tilde{I}$, and it can be divided into four branches 
\bea
\tilde{G}=G^{C^{+}P^{+}}\sqcup G^{C^{+}P^{-}} \sqcup G^{C^{-}P^{+}} \sqcup G^{C^{-}P^{-}}\,,
\eea
 with 
 \begin{equation}
 \left\{
 \begin{array}{rll}
G^{^{C^{+}P^{+}}}\!\!\!\!\!\!&\equiv \{ g_{_{\{C^{+}P^{+}\}}}:= g &| g\in G  \}
 \\
 G^{^{C^{+}P^{-}}}\!\!\!\!\!\!&\equiv \{ g_{_{\{C^{+}P^{-}\}}}:= g\mathcal{P} &| g\in G\}
 \\
 G^{^{C^{-}P^{+}}}\!\!\!\!\!\!&\equiv \{ g_{_{\{C^{-}P^{+}\}}}:= g\mathcal{C} &| g\in G\}
 \\
 G^{^{C^{-}P^{-}}}\!\!\!\!\!\!&\equiv \{ g_{_{\{C^{-}P^{-}\}}}:= g\mathcal{CP} &| g\in G\}
 \end{array}
 \right..
\end{equation}
We also define $SO(4)$ branches as  $G^{^{C^{+}P^{+}}}\sqcup G^{^{C^{-}P^{-}}}$ and $O_{-}(4)$ branches as  $G^{^{C^{-}P^{+}}}\sqcup G^{^{C^{+}P^{-}}}$. 
Different branches are connected by group element $\mathcal{C}$ and $\mathcal{P}$. 

Since we have added $\mathcal{C}$ and $\mathcal{P}$ as new group elements into the group $G$, the original representation $\bf R$ of $G^{C^{+}P^{+}}$ should be also a representation of $\tilde{G}$. It's equivalent to say that the action (defined in field theory) of $\mathcal{C}\mathcal{P}$ do not change the representation
\bea
\bf R=\bf R^{\mathcal{C}} =\bf R^{\mathcal{P}}=\bf R^{\mathcal{CP}}.
\eea
 This is nothing but the fact that the anti-particles and the left and right hand particles should be contained in the theory.

By extending the group $G$ to $\tilde{G}$, the Hilbert series in Eq.~\ref{eq:finalHS} becomes
\bea
\mathcal{H}^{CP}  (D, \phi )=\frac{1}{4}\sum_{C^{\pm}P^{\pm}} \mathcal{H}^{C^{\pm}P^{\pm}} (D, \phi ),
\eea
with the ones in the four branches
\bea
\mathcal{H}^{C^{\pm}P^{\pm}}(D, \phi ) \equiv
\int_{G} d\mu(g)
\left(
\sum_{C^{\pm}P^{\pm}} 
\frac{\text{PE}(\phi \chi_{_{R_{\phi}}}(D,g_{_{\{C^{\pm}P^{\pm}\}}}))}{P(D,g_{_{\{C^{\pm}P^{\pm}\}}}))}
\right).
\label{eq:HSCPfour}
\eea
The $\mathcal{H}^{C^{+}P^{+}}$ is exact what we have gotten in Eq.~\ref{eq:finalHS} for single bosonic particle case. Although we have moved the integral onto the $C^{+}P^{+}$ branch by using the invariance of the Haar measure, the parametrization for the integrand in the integration in different branches are different. We use $a$ ($a_{-}$) and $b$ ($b_{-}$) to parametrize the integral on $SO(4)$ ($O_{-}(4)$) and $C^{+}$ ($C^{-}$) branches. The parametrization for different branches are listed in Table~\ref{CPtable1}.


\begin{table}[htb]
\centering
\resizebox{\textwidth}{!}{
\begin{tabular}{|c|clcccl|}
\hline
Group Branch                                      & \multicolumn{4}{c|}{$SO(4)$}                                                                        & \multicolumn{2}{c|}{$O_{-}(4)$}\\ \hline
integral variable                                 &
\multicolumn{4}{c|}{$a_{+}=a=(a_{1},a_{2})$}
                                                  &

\multicolumn{2}{c|}{$a_{-}=a_{1}$}\\ \hline
reparametrization                                &
\multicolumn{4}{c|}{$\bar{a}_{+}=a$}
                                                  &

\multicolumn{2}{c|}{$\bar{a}_{-}=(a_{1},1)$}\\ \hline
Haar measure                                      &
\multicolumn{4}{c|}{$d\mu_{SO(4)}(a)$}                                                                      & \multicolumn{2}{c|}{$d\mu_{Sp(2)}(a_{-})$}
\\ \hline
 \multicolumn{7}{|c|}{}
\\
\hline
Group Branch                                      & \multicolumn{4}{c|}{$U(1)$}                                                                        & \multicolumn{2}{c|}{$U_{-}(1)$}\\ \hline
integral variable                                 &
\multicolumn{4}{c|}{$x_{+}=x$}
                                                  &

\multicolumn{2}{c|}{$x_{-}=x$}\\ \hline
reparametrization                                &
\multicolumn{4}{c|}{$\bar{x}_{+}=x$}
                                                  &

\multicolumn{2}{c|}{$\bar{x}_{-}=x$}\\ \hline
Haar measure                                      &
\multicolumn{4}{c|}{$d\mu_{U(1)}(x)$}                                                                      & \multicolumn{2}{c|}{$d\mu_{U(1)}(x)$}
\\ \hline
 \multicolumn{7}{|c|}{}
\\
\hline
Group Branch                                      & \multicolumn{4}{c|}{$SU(2)$}                                                                       & \multicolumn{2}{c|}{$SU_{-}(2)$}                                                                              \\ \hline
integral variable                                 &
\multicolumn{4}{c|}{$y_{+}=y$}
                                                  &

\multicolumn{2}{c|}{$y_{-}=y$}\\ \hline
reparametrization                                &
\multicolumn{4}{c|}{$\bar{y}_{+}=y$}
                                                  &

\multicolumn{2}{c|}{$\bar{y}_{-}=y$}\\ \hline
Haar measure                                      & \multicolumn{4}{c|}{$d\mu_{SU(2)}(y)$}                                                                 & \multicolumn{2}{c|}{$d\mu_{SU(2)}(y)$}                                                                        \\ \hline
 \multicolumn{7}{|c|}{}
\\
\hline
Group Branch                                      & \multicolumn{2}{c|}{$SU(N)$}                                     & \multicolumn{2}{c|}{$SU_{-}(N=2k)$}                                   & \multicolumn{2}{c|}{$SU_{-}(N=2k+1)$}                                   \\ \hline
integral variable                                 &
\multicolumn{2}{c|}{$z=(z_{1},...,z_{N})$}
                                                  &
\multicolumn{2}{c|}{$z_{-}=(z_{1},...,z_{k})$}
                                                  &
\multicolumn{2}{c|}{$z_{-}=(z_{1},...,z_{k})$}\\ \hline
reparametrization                                 &
\multicolumn{2}{c|}{$\bar{z}_{+}=z$}
                                                  &
\multicolumn{2}{c|}{$\bar{z}_{-}=(\sqrt{z_{1}},...,\sqrt{z_{k}},1/\sqrt{z_{k}},...,1/\sqrt{z_{1}})$}
                                                  &
\multicolumn{2}{c|}{$\bar{z}_{-}=(\sqrt{z_{1}},...,\sqrt{z_{k}},1,1/\sqrt{z_{k}},...,1/\sqrt{z_{1}})$}\\ \hline
Haar measure                                      & \multicolumn{2}{c|}{$d\mu_{SU(N)}(z)$}                               & \multicolumn{2}{c|}{$d\mu_{SO(2k+1)}(z_{-})$}                     & \multicolumn{2}{c|}{$d\mu_{Sp(2k)}(z_{-})$}                         \\ \hline
\end{tabular}
}
\caption{The parameter (of integration and characters) in different branches of $\tilde{G}$ is given in this table. Here $z_{i}$ in $SU(N)$ satisfy $z_{1}...z_{N}=1$ hence the degree of freedom is $N-1$ ($N\geq3$).}
\label{CPtable1}
\end{table}

\begin{table}[ht]
\centering
\resizebox{\textwidth}{!}{
\begin{tabular}{|c|clcccl|}
\hline
Group Branch                                      & \multicolumn{4}{c|}{$SO(4)$}                                                                        & \multicolumn{2}{c|}{$SO_{-}(4)$}\\ \hline

$tr(g^{2n-1})$\; for\;$(j_{1},j_{1})$              & \multicolumn{4}{c|}{$\chi_{(j_{1},j_{1})}^{SO(4)}(a^{2n-1})$}                                                 & \multicolumn{2}{c|}{$\eta_{P}\chi_{j_{1}+j_{2}}^{Sp(2)}(a_{-}^{2n-1})$}                                                    \\ \hline
$tr(g^{2n-1})$\; for\;$j_{1}\neq j_{2}$ & \multicolumn{4}{c|}{$\chi_{(j_{1},j_{2})}^{SO(4)}(a^{2n-1})$}  & \multicolumn{2}{c|}{0}                                                                                        \\ \hline
 \multicolumn{7}{|c|}{}
\\
\hline
Group Branch                                      & \multicolumn{4}{c|}{$U(1)$}                                                                        & \multicolumn{2}{c|}{$U_{-}(1)$}\\ \hline

$tr(g^{2n-1})$\; for singlet              & \multicolumn{4}{c|}{$1$}                                                 & \multicolumn{2}{c|}{$\eta_{C}$}                                                    \\ \hline
$tr(g^{2n-1})$\; for\;$q$\; rep & \multicolumn{4}{c|}{$\chi_{q}^{U(1)}(x^{2n-1})$}  & \multicolumn{2}{c|}{$0$}                                                                                        \\ \hline
 \multicolumn{7}{|c|}{}
\\
\hline
Group Branch                                      & \multicolumn{4}{c|}{$SU(2)$}                                                                       & \multicolumn{2}{c|}{$SU_{-}(2)$}                                                                              \\ \hline

$tr(g^{2n-1})$\; for\;singlet\; rep                        & \multicolumn{4}{c|}{1}                                                                                 & \multicolumn{2}{c|}{$\eta_{C}$}                                                                                        \\ \hline
$tr(g^{2n-1})$\; for\; adjoint\; rep                 & \multicolumn{4}{c|}{$\chi_{\text{ad}}^{SU(N)}(y^{2n-1})$}                                                            & \multicolumn{2}{c|}{$\eta_{C} \chi_{\text{ad}}^{SU(2)}(y^{2n-1})$}                                                        \\ \hline
$tr(g^{2n-1})$\; for\; fund\; rep           &\multicolumn{4}{c|}{$\chi_{\text{fund}}^{SU(2)}(y^{2n-1})$}                        & \multicolumn{2}{c|}{0*}                 \\ \hline
 \multicolumn{7}{|c|}{}
\\
\hline
Group Branch                                      & \multicolumn{2}{c|}{$SU(N)$}                                     & \multicolumn{2}{c|}{$SU_{-}(N=2k)$}                                   & \multicolumn{2}{c|}{$SU_{-}(N=2k+1)$}                                   \\ \hline

$tr(g^{2n-1})$\; for\; singlet\; rep                 & \multicolumn{2}{c|}{1}                                               & \multicolumn{2}{c|}{$\eta_{C}$}                                                & \multicolumn{2}{c|}{$\eta_{C}$}                                                  \\ \hline

$tr(g^{2n-1})$\; for\; adjoint\; rep                 & \multicolumn{2}{c|}{$\chi_{\text{ad}}^{SU(N)}(z^{2n-1})$}                      & \multicolumn{2}{c|}{$\eta_{C} \chi_{\text{fund}}^{SO(2k+1)}(z_{-}^{2n-1})$} & \multicolumn{2}{c|}{$\eta_{C} \chi_{\text{fund}}^{Sp(2k)}(z_{-}^{2n-1})$}     \\ \hline
$tr(g^{2n-1})$\; for\; fund(anti)\; rep                 & \multicolumn{2}{c|}{$\chi_{\text{fund(anti)}}^{SU(N)}(z^{2n-1})$}                      & \multicolumn{2}{c|}{0} & \multicolumn{2}{c|}{0}     \\ \hline
\end{tabular}
}
\caption{Here we list the character building blocks of $\chi_{R_{\phi}}^{\text{branch}}$ in Eq.~\ref{eq:generating1} Eq.~\ref{eq:generating2}. For any determinant irrep $\bf{r}$ the character $\chi_{\bf{r}}$ in certain branch is obtained by multiplying the corresponding character in this table altogether. The * after character indicates the possibility that the anti-particle is the same as the particle. This possibility comes from the fact that SU(2) group do not really have an outer automorphism, and the anti-fundamental representation is identical with fundamental representation. If this happens, the character in $C^{-}$ branch should be $\eta_{C}\chi_{\text{anti}}^{SU(2)}(y^{2n-1})$.}
\label{CPtable2}
\end{table}


The factor P in Eq.~\ref{eq:kinematical factor} do not change in $SO(4)$ branches. In $O_{-}(4)$ branches it is parametrized as
\begin{equation}
    P(D,g_{_{\{O_{-}(4)\}}}(a_{-}))=\frac{1}{(1-Da_{1})(1-Da_{1}^{-1})(1-D^2)}
\end{equation}
with the parameter $a_{-}=a_1$ in Table~\ref{CPtable1}. This corresponds to the fact that there exist a similar matrix $X$ such that $O_{-}(4)\ni g=X\text{diag}(a_{1},a_{1}^{-1},1,-1)X^{-1}$. 

The sum in the PE function in Eq.~\ref{eq:HSCPfour} splits into the odd-power and even-power terms~\cite{Henning:2017fpj,Graf:2020yxt}
\begin{equation}
    \ln \text{PE}(\phi \chi_{_{R_{\phi}}}(D,g_{_{\{C^{\pm}P^{\pm}\}}}))=
    \sum_{n=1}\left(\frac{\phi^{2n}}{2n}\chi^{C^{+}P^{+}}_{_{R_{\phi}}}(D^{2n},\bar{a}_{\pm}^{2n},\bar{b}_{\pm}^{2n})+\frac{\phi^{2n-1}}{2n-1}\chi^{C^{\pm}P^{\pm}}_{_{R_{\phi}}}(D^{2n-1},a_{\pm}^{2n-1},b_{\pm}^{2n-1})\right).
    \label{eq:generating1}
\end{equation}
The even-power terms are related to the characters Eq.~\ref{eq:character} in the $C^{+}P^{+}$ branch with only a change of parameters (also listed in Table~\ref{CPtable1}),
\bea
a
\rightarrow
\left\{
\begin{array}{ll}
    \bar a_{+}=a &\text{in }  SO(4) \text{ branches}\\
    \bar a_{-}=\bar{a} &\text{in }  O_{-}(4) \text{ branches}
\end{array}
\right.
\,,\,
b
\rightarrow
\left\{
\begin{array}{ll}
    \bar b_{+}=b &  C^{+}\; \text{branches} \\
    \bar b_{-}=\bar{b} &  C^{-}\;\text{branches}
\end{array}
\right..
\eea
The odd-power terms are totally different from the terms in the $C^{+}P^{+}$ branch and we list the main results in Table~\ref{CPtable2}.
When fermionic field is involved, the PE should be modified as
\begin{equation}
    \ln \text{PE}_{f}(\phi \chi_{_{R_{\phi}}}(D,g_{_{\{C^{\pm}P^{\pm}\}}}))=-
    \sum_{n=1}\left(\frac{\phi^{2n}}{2n}\chi^{C^{+}P^{+}}_{_{R_{\phi}}}(D^{2n},\bar{a}_{\pm}^{2n},\bar{b}_{\pm}^{2n})-\frac{\phi^{2n-1}}{2n-1}\chi^{C^{\pm}P^{\pm}}_{_{R_{\phi}}}(D^{2n-1},a_{\pm}^{2n-1},b_{\pm}^{2n-1})\right).
    \label{eq:generating2}
\end{equation}
The specific forms of the $\bar{a}_{\pm}$ and $\bar{b}_{\pm}$ presented in Ref.~\cite{Henning:2017fpj,Graf:2020yxt} are valid for the $SO(4)$ and general $SU(N)$ group with $N \ge 3$, respectively. For the $U(1)$ and $SU(2)$ groups, $\bar{b}_{\pm}$ needs to be determined. We have derived them and listed them in Table~\ref{CPtable1}.



In the following, we show how to derive the character PE in the $U(1)$ group, which can be similarly applied to the $SU(2)$ case. It is convenient to consider charge conjugation of a $U(1)$ gauge group. Since the Lorentz group is not involved, we do not need to consider parity $\mathcal{P}$. Assume that $\phi \rightarrow  e^{iq\theta}\phi$ be in the representation $q$ of the group $U(1)$ and its anti-particle $\phi^{\dagger} \rightarrow  e^{-iq\theta}\phi^{\dagger}$. The symmetry group now becomes
\begin{equation}
    \tilde{U}(1)=\{x,x\mathcal{C}|x=e^{i\theta},\theta\in [0,2\pi)\}.
\end{equation}

Since the gauge group $\tilde{U}(1)$ is divided into the $C^{+}$ and $C^{-}$ branches, the Hilbert series is obtained by integrating the character PE on these two branches. Here we use one formal parameter $\phi$ in Hilbert series to count the total degree of $\phi$ and $\phi^{\dagger}$ because these two representations of $U(1)$ are connected by $\mathcal{C}$. Given the group element $g_{\{C^{+}\}} = g\in G^{C^{+}}$,  $g_{\{C^{-}\}}$ is a group element in the $C^{-}$ branch which can be written as $g_{\{C^{-}\}}=g\mathcal{C}$. The charge conjugation $\mathcal{C}$ simply turns $\phi$ and $\phi^{\dagger}$ into each other, so the representation matrices for the group elements $g$ and $ g\mathcal{C}$ read
\bea
\left(\begin{array}{cc}
        x &  \\
         & x^{-1}
    \end{array}
    \right), \quad
   \left(\begin{array}{cc}
         & x \\
        x^{-1} &
    \end{array}
    \right).
\eea

Given the group elements and their representations, the character PEs on the $C^{\pm}$ branches are obtained as
\begin{equation}
\begin{split}
    \text{PE}(\phi\chi_{_{R_{\phi}}}(g_{C\pm}))&=\exp\left(\sum_{n=1}\frac{\phi^{n}\chi_{_{R_{\phi}}}(g_{\{C\pm\}})^{n}}{n}\right) 
    = \left\{
\begin{array}{ll}
    \exp\left(\sum_{n=1}\frac{\phi^{n}(\chi_{q}(x^n)+\chi_{-q}(x^n))}{n}\right) &\text{in } C_+ \text{ branch}\\
    \exp\left(\sum_{n=1}\frac{\phi^{2n}(\chi_{q}(1)+\chi_{-q}(1))}{2n}\right) &\text{in }  C_- \text{ branch}
\end{array}
\right.
\end{split}
\label{eq:PEofC}
\end{equation}
%
If $\phi$ is a fermionic particle, neglecting the Lorentz structure for now, the character PE is 
\bea
    \text{PE}_{f}(\phi\chi_{_{R_{\phi}}}(g_{C\pm}))    
    = \left\{
\begin{array}{ll}
    \exp\left( -\sum_{n=1}\frac{(-\phi)^{n}(\chi_{q}(x^n)+\chi_{-q}(x^n))}{n}\right) &\text{in } C_+ \text{ branch}\\
    \exp\left( -\sum_{n=1}\frac{\phi^{2n}(\chi_{q}(1)+\chi_{-q}(1))}{2n}\right) &\text{in }  C_- \text{ branch}
\end{array}
\right.
\eea
The total HS is thus
\begin{equation}
    H(\phi)=\frac{1}{2}\int \frac{dx}{2\pi ix}\left(
    \text{PE}(\phi\chi_{_{R_{\phi}}}(g_{C+}))+
    \text{PE}(\phi\chi_{_{R_{\phi}}}(g_{C-}))
    \right).
\end{equation}
which gives the number of invariants under the $U(1)\rtimes \mathcal{C}$ symmetry. There are two properties in this example which are also true for $SO(4)$ and $SU(N)$ groups. Firstly the odd power in Eq.~\ref{eq:PEofC} vanishes because $\mathcal{C}$ changes the $\phi$ into $\phi^{\dagger}$. Secondly the even power terms in Eq.~\ref{eq:PEofC} on the $C^{-}$ branch is related to the character $\chi_{q}(x)$ on $C^{+}$ with a re-parametrization $x\rightarrow 1$.


\section{Effective Operators in Higgs EFT}
\label{sec:HEFT}



\subsection{Building Blocks and LO Lagrangian}

In the SM, the electroweak symmetry spontaneously breaking pattern is
\begin{equation}
    \mathcal{G} = SU(2)_L\times SU(2)_R \rightarrow \mathcal{H} = SU(2)_V\,,
\end{equation}
which can be characterised by the CCWZ formalism~\cite{Coleman:1969sm,Callan:1969sn}. Promoting the symmetry breaking generators into the dynamical fields, the NGBs can be parametrized into the exponential unitary matrix
\begin{equation}
    \bfu(x) = \exp(-i\frac{\pi(x)\cdot \sigma}{f})\,,
\end{equation}
where $\sigma$'s are the generators for the coset symmetry, $\pi(x)$'s are the 3 NGB fields
\bea
\label{eq:ngbs}
\pi(x)=(\pi_1(x),\pi_2(x),\pi_3(x))\,,
\eea
and $f$ is the symmetry breaking scale, served as the dimension compensation to make the quantity in the exponential dimensionless.

In the chiral Lagrangian framework, the SM Higgs doublet $H(x)$ can be reparametrized as \bea
\Sigma(x) = (\tilde{H}(x),H(x)) =
\frac{h(x)+v}{\sqrt{2}} \bfu(x) \,,
\eea
where $\tilde{H}=i\sigma_2 H^*$. Both the field $\Sigma(x)$ and $\bfu(x)$ transform under the group $\mathcal{G}$ as the  bi-doublet
\bea
     \Sigma(x) \rightarrow \mathfrak{g}_L \Sigma(x) \mathfrak{g}^\dagger_R \quad
     \bfu \rightarrow \mathfrak{g}_L\bfu \mathfrak{g}^\dagger_R\,,\quad (\mathfrak{g}_L,\mathfrak{g}_R)\in \mathcal{G}=SU(2)_L \times SU(2)_R\,,
\eea
and the Higgs field now transforms as the singlet $h(x)$ under the group.

For the fermions $\psi_{L/R}=P_{L/R}\psi$ in the SM, they can be cooperated to the symmetry group $\mathcal{G}$ by the redefinition,
\begin{align}
Q_L &= \left(\begin{array}{c}u_L\\d_L\end{array}\right) \quad \rightarrow \quad \mathfrak{g}_{L} Q_L\,, \quad \quad
Q_R = \left(\begin{array}{c}u_R\\d_R\end{array}\right) \quad \rightarrow \quad \mathfrak{g}_{R} Q_R\,, \\ 
L_L &= \left(\begin{array}{c}\nu_L\\e_L\end{array}\right) \quad \rightarrow \quad \mathfrak{g}_{L} L_L\,, \quad \quad
L_R = \left(\begin{array}{c}\nu_R\\e_R\end{array}\right) \quad \rightarrow \quad \mathfrak{g}_{R} L_R\,,
\end{align}
where $\mathfrak{g}_L\in SU(2)_L\,,\mathfrak{g}_R\in SU(2)_R$. Since the right-handed spinors are collected to form the $SU(2)_R$ doublets, the hyper charge symmetry group $U(1)_Y$ are prompted to the $U(1)_X$, where $X=(B-L)/2$ is the half of the baryon number minus the lepton number. In the Higgs EFT, typically right-handed neutrinos are absent in the building blocks, while with the right-handed neutrinos included, it is extended to be the HEFT with light sterile neutrinos, the so-called $\nu$HEFT, the broken theory of the SMEFT with sterile neutrinos ($\nu$SMEFT)~\cite{Aparici:2009fh,delAguila:2008ir,Bhattacharya:2015vja,Liao:2016qyd,Li:2021tsq}. In the following, we would like to keep the sterile neutrinos in the right-handed doublet $L_R$, and so it is more convenient to compare the results in the Hilbert series with the ones in the Young tensor method.

In terms of the fermion doublets above, the Yukawa interactions in the SM takes the form that
\begin{equation}
    	v \overline{\psi_L} \bfu {\mathcal Y}^\psi_R  \psi_R + h.c. \quad {\textrm{with}} \quad {\mathcal Y}^\psi_R  \quad \longrightarrow \quad \mathfrak{g}_{R} {\mathcal Y}^\psi_R \mathfrak{g}^\dagger_{R}\,,\quad \mathfrak{g}_R\in SU(2)_R\,,
\end{equation}
where $\psi=Q,L$, which is not invariant under the custodial symmetry $SU(2)_V$ unless the Yukawa coupling matrix $\mathcal{Y}^\psi_R$ is proportional to the identity matrix. Generally, the coupling matrix takes the form that
\begin{equation}
    \mathcal{Y}^Q_R = \frac{1}{2}(y_u+y_d)+\frac{1}{2}(y_u-y_d)\sigma_3\,,\quad \mathcal{Y}^L_R = \frac{1}{2}(y_\nu+y_e)+\frac{1}{2}(y_\nu-y_e)\sigma_3\,,
\end{equation}
where $y_u,y_d,y_\nu,y_e$ are the Yukawa coupling constants in the SM, and $y_\nu=0$ if there is no right-handed neutrino. Taking the $\mathcal{T}_R=\sigma_3/2$ as the spurion, this custodial symmetry breaking property can be characterised by it entirely.

Besides, promoting the normal derivative to the covariant one generates 3 gauge fields, $G\,,W\,,B$, corresponding to the color, electroweak and hyper charge symmetry respectively. In the construction of the effective operators, it is convenient to adopt the chiral components of their field-strength tensors as fundamental building blocks,
\begin{align}
    {X_L}_{\mu\nu} = \frac{1}{2}(X_{\mu\nu}-i\tilde{X}_{\mu\nu})\,,\quad {X_R}_{\mu\nu} = \frac{1}{2}(X_{\mu\nu}+i\tilde{X}_{\mu\nu})\,,
\end{align}
where $\tilde{X}_{\mu\nu}=\epsilon_{\mu\nu\rho\lambda}X^{\rho\lambda}$ and $X=G,W,B$.
Furthermore, it is helpful to redefine these building blocks which is of the representations of $\mathcal{G}$ to be the form transforming solely under the group $SU(2)_L$~\cite{Buchalla:2012qq,Alonso:2012px,Brivio:2013pma,Buchalla:2013rka,Gavela:2014vra,Brivio:2016fzo,Merlo:2016prs,Krause:2018cwe},
\begin{align}
	\bfv_\mu(x) = i\bfu(x) D_\mu \bfu(x)^\dagger \quad  &\longrightarrow  \quad \mathfrak{g}_L \bfv_\mu \mathfrak{g}_L^\dagger\,, \label{eq:vmu} \notag\\
	\mathbf{T} = \bfu \mathcal{T}_{R} \bfu^{\dagger} \quad &\longrightarrow  \quad \mathfrak{g}_{L} \mathbf{T} \mathfrak{g}_{L}^\dagger\,,\notag\\
	 \psi_R = \bfu \psi_{R} \quad &\longrightarrow  \quad \mathfrak{g}_{L} \psi_{R}\,,
\end{align}
There are other redefinitions of the building blocks such as in Ref~\cite{Pich:2015kwa,Pich:2016lew,Pich:2018ltt}, and different choices give the same operators set. 

In terms of these building blocks, the leading-order (LO) Lagrangian is known as
\begin{align}
\mathcal{L}_{\text{LO}} =&-\frac{1}{4} \left( G^a_{\mu\nu}G^{a\mu\nu}\right) - \frac{1}{4}\lra{\hat{W}_{\mu\nu} \hat{W}^{\mu\nu}} - \frac{1}{4}B_{\mu\nu}B^{\mu\nu} -\frac{g_s^2}{16\pi^2}\theta_s \left( G^a_{\mu\nu}\tilde{G}^{a\mu\nu} \right) \notag \\
& + \frac{1}{2}\partial_\mu h \partial^\mu h -V(h) -\frac{v^2}{4}\lra{\bfv_\mu \bfv^\mu}\mathcal{F}_C(h) - \frac{v^{2}}{4} \lra{\bft \bfv_{\mu}}\lra{\bft \bfv^{\mu}} \mathcal{F}_{T}(h) \notag \\
& +i\bar{Q}_L\slashed{D} Q_L + i\bar{Q}_R\slashed{D}Q_R + i\bar{L}_L\slashed{D}L_L + i\bar{L}_R \slashed{D}L_R \notag \\
& -\frac{v}{\sqrt{2}}(\bar{Q}_L \bfu \mathcal{Y}^Q_R(h) Q_R) + h.c.) -\frac{v}{\sqrt{2}}(\bar{L}_L \bfu \mathcal{Y}^L_R(h)L_R + h.c.)\,,\label{eq:lol}
\end{align}
where $\lra{\dots}$ represents the $SU(2)_L$ trace. $\mathcal{F}_C(h)$ and $\mathcal{F}_T(h)$ appearing in the second lines are dimensionless polynomials of Higgs $h$. 

In analogy with the ChPT, we adopt the chiral dimension $d_\chi$~\cite{Buchalla:2013eza,Gavela:2016bzc} to organise the HEFT effective operators. We define the building block $\bfv_\mu$ is of chiral dimension 1, and the NGBs matrix $\bfu$ is chiral-dimensionless, thus the chiral dimension of the normal or covariant derivative is 1. As a result, the LO Lagrangian in Eq.~\ref{eq:lol} is of chiral dimension 2, thus the chiral dimension of all other building blocks are determined. We summarise the building blocks of HEFT, together with their chiral dimension and group representations in Table~\ref{tab:buildingblocks}, and the explicit expression of the group characters are presented in Tab.~\ref{tab:char}. In some literature, the spurion $\bft$ is also be set of $d_\chi=1$, the effective operators at specific dimension is different under these 2 versions of power-counting scheme, while the complete set of operators should be the same.

\begin{table}[]
    \centering
    \resizebox{\textwidth}{!}{
    \begin{tabular}{c|c|c|c|c|c|c}
    \hline
    $\Phi$ & $d_\chi$ & Lorentz group & $SU(2)_L$ & $SU(3)_C$ & $U(1)_X$ & Character $\chi_\Phi$ \\
    \hline
    \hline
    $\bfv_\mu$ & 1 & $(1,0)$ & $\mathbf{3}$ & $\mathbf{0}$ & 0 & $\chi_{R_\bfv}(D,a)\chi_{SU(2)}^{\mathbf{3}}(y)$ \\
    $Q_{L/R}$ & 1 & $(\frac{1}{2},0)/(0,\frac{1}{2})$ & $\mathbf{2}$ & $\mathbf{3}$ & $\frac{1}{6}$ & $\chi_{R_{\psi_{L/R}}}(D,a)\chi^{\frac{1}{6}}(x)\chi^{\mathbf{2}}_{SU(2)}(y)\chi^{\mathbf{3}}_{SU(3)}(z_1,z_2)$ \\
    $Q^\dagger_{L/R}$ & 1 & $(0,\frac{1}{2})/(\frac{1}{2},0)$ & $\mathbf{2}$ & $\overline{\mathbf{3}}$ & $-\frac{1}{6}$ & $\chi_{R_{\psi_{R/L}}}(D,a)\chi^{-\frac{1}{6}}(x)\chi^{\mathbf{2}}_{SU(2)}(y)\chi^{\overline{\mathbf{3}}}_{SU(3)}(z_1,z_2)$ \\
    $L_{L/R}$ & 1 & $(\frac{1}{2},0)/(0,\frac{1}{2})$ & $\mathbf{2}$ & $\mathbf{0}$ & $-\frac{1}{2}$ & $\chi_{R_{\psi_{L/R}}}(D,a)\chi^{-\frac{1}{2}}(x)\chi^{\mathbf{2}}_{SU(2)}(y)$ \\
    $L^\dagger_{L/R}$ & 1 & $(0,\frac{1}{2})/(\frac{1}{2},0)$ & $\mathbf{2}$ & $\mathbf{0}$ & $\frac{1}{2}$ & $\chi_{R_{\psi_{R/L}}}(D,a)\chi^{\frac{1}{2}}(x)\chi^{\mathbf{2}}_{SU(2)}(y)$ \\
    $W$ & 2 & $(1,1)\oplus(1,-1)$ & $\mathbf{3}$ & $\mathbf{0}$ & 0 & $(\chi_{R_{F_L}}(D,a)+\chi_{R_{F_R}}(D,a))\chi^{\mathbf{3}}_{SU(2)}(y)$\\
    $G$ & 2 & $(1,1)\oplus(1,-1)$ & $\mathbf{0}$ & $\mathbf{8}$ & 0 & $(\chi_{R_{F_L}}(D,a)+\chi_{R_{F_R}}(D,a))\chi^{\mathbf{8}}_{SU(3)}(z_1,z_2)$\\
    $B$ & 2 & $(1,1)\oplus(1,-1)$ & $\mathbf{0}$ & $\mathbf{0}$ & 0 & $(\chi_{R_{F_L}}(D,a)+\chi_{R_{F_R}}(D,a))$ \\
    $h$ & 0 & $(0,0)$ & $\mathbf{0}$ & $\mathbf{0}$ & 0 & $\chi_{R_\varphi}(D,a)$\\
    $\bft$ & 0 & $(0,0)$ & $\mathbf{3}$ & $\mathbf{0}$ & 0 & $\chi_{R_\bft}(D,a)\chi^{\mathbf{3}}_{SU(2)}(y)$ \\
    \hline
    \end{tabular}
    }
    \caption{The building blocks $\Phi$ of the HEFT, where $d_\chi$ is the chiral dimension, $SU(3)_C\times SU(2)_L$ is the non-Abelien part of the SM gauge group, and $U(1)_X$ is the counterpart of the Abelien part $U(1)_Y$ of the SM gauge group in the HEFT. The representations of the Lorentz group are labeled by $(j_1,j_2)$.}
    \label{tab:buildingblocks}
\end{table}

\begin{table}[]
    \centering
    \resizebox{\textwidth}{!}{
    \begin{tabular}{c|c|c|c}
    \hline
Group & Representation & Character & Haar measure \\
\hline
\hline
\multirow{6}{*}{Lorentz} & $(0,0)$ & 1 &  \multirow{5}{*}{$\frac{1}{(2\pi i)^2}\frac{1}{4a_1a_2}(1-\frac{a_1}{a_2})(1-\frac{a_2}{a_1})\times$}\\
& $(\frac{1}{2},\frac{1}{2})$ & $a_1+\frac{1}{a_1}+a_2+\frac{1}{a_2}$ & \multirow{5}{*}{$(1-a_1a_2)(1-\frac{1}{a_1a_2})\text{d}a_1\text{d}a_2$}\\
& $(\frac{1}{2},0)$ & $\sqrt{a_1a_2}+\sqrt{\frac{1}{a_1a_2}}$ & \\
& $(0,\frac{1}{2})$ & $\sqrt{\frac{a_1}{a_2}}+\sqrt{\frac{a_2}{a_1}}$ & \\
& $(1,0)$ & $a_1a_2+1+\frac{1}{a_1a_2}$ & \\
& $(0,1)$ & $\frac{a_1}{a_2}+1+\frac{a_2}{a_1}$ & \\
\hline
$U(1)$ & $Q$ & $x^Q$ & $\frac{1}{2\pi i}\frac{1}{x}\text{d}x$ \\
\hline
\multirow{3}{*}{$SU(2)$} 
& \textbf{0} & 1 & \multirow{3}{*}{$\frac{1}{2\pi i}\frac{1}{2y}(1-y^2)(1-y^{-2})\text{d}y$} \\
& \textbf{2} & $y+\frac{1}{y}$ &  \\
& \textbf{3} & $y^2+1+\frac{1}{y^2}$ & \\
\hline
\multirow{4}{*}{$SU(3)$} 
& \textbf{0} & 1 & \multirow{3}{*}{$\frac{1}{(2\pi i)^2}\frac{1}{6z_1z_2}(1-z_1^2z_2)(1-\frac{1}{z_1z_2^2})\times$} \\
& \textbf{3} & $z_1+\frac{1}{z_1z_2}+z_2$ & \multirow{3}{*}{$(1-\frac{z_2}{z_1})(1-\frac{z_1}{z_2})(1-z_1z_2^2)(1-\frac{1}{z_1^2z_2})\text{d}z_1\text{d}z_2$}\\
& $\bar{\textbf{3}}$ & $\frac{1}{z_1}+z_1z_2+\frac{1}{z_2}$ & \\
& \textbf{8} & $\frac{z_1}{z_2}+\frac{1}{z_1z_2^2}+z_1^2z_2+2+z_1z_2^2+\frac{1}{z_1^2z_2}+\frac{z_2}{z_1}$ \\
\hline
$Sp(2)$ & fundamental & $y+\frac{1}{y}$ & $\frac{1}{2\pi i}\frac{1}{2y}(1-y^2)(1-y^{-2})\text{d}y$ \\
\hline
$SO(3)$ & fundamental & $y+1+\frac{1}{y}$ & $\frac{1}{2\pi i}\frac{1}{2y}(1-y)(1-\frac{1}{y})\text{d}y$ \\
\hline
    \end{tabular}
    }
    \caption{The characters and the Haar measures of the Lorentz and some classical groups used in this work.}
    \label{tab:char}
\end{table}

\subsection{Higher-order Lagrangian up to chiral dimension 10}
The general form of the HEFT Lagrangian takes the form that
\begin{equation}
    \mathcal{L}=\mathcal{L}^{\textrm LO}_{d_\chi =2} + \mathcal{L}^{\textrm NLO}_{d_\chi =3,4} + \mathcal{L}^{\textrm NNLO}_{d_\chi =5,6} + \sum_{d_\chi>6}^\infty \mathcal{L}_{d_\chi}\,,
\end{equation}
where we have identified the LO Lagrangian to be of chiral dimension 2.
Beyond the LO Lagrangian, the NLO Lagrangian has been discussed in Ref~\cite{Alonso:2012px,Buchalla:2012qq,Brivio:2013pma,Buchalla:2013rka,Gavela:2014vra,Pich:2015kwa,Pich:2016lew,Brivio:2016fzo,Merlo:2016prs,Krause:2018cwe,Pich:2018ltt}, and the complete and independent operator set is first obtained in Ref.~\cite{Sun:2022ssa}. For the NNLO operators, the complete and the independent set are constructed in Ref~\cite{Sun:2022snw} for the first time. Since of the special building blocks such as $\bfv_\mu,\bft\,,h$ in the HEFT, extra manipulation is needed in the constructions of the higher-dimension operators,
\begin{itemize}
    \item The NGBs from the symmetry breaking satisfy the Adler zero condition~\cite{Adler:1964um,Adler:1965ga}, which means that there should be at least one derivative applying on them, thus the corresponding building block in the HEFT is taken as $\bfv_\mu$. To identify the SPM of such building block, we notice that the linear expansion of $\bfv_\mu$ has the leading term
\begin{equation}
    \bfv_\mu(x)=i\bfu(x) D_\mu \bfu(x)^\dagger\sim D_\mu\pi(x)\,, 
\end{equation}
where $\pi(x)$ is the NGB in Eq.~\ref{eq:ngbs}, thus the SPM of $\bfv_\mu$ is just the one of scalar field in Eq.~\ref{eq:rphi} with the top sub-module removed, 
\bea
R_{\bfv} = \left(
\begin{array}{c}
D_{\mu}\pi\\
D_{\{\mu_{1}}D_{\mu_{2}\}}\pi \\
\vdots
\end{array}
\right)
\,.
\eea
whose corresponding character is 
\begin{equation}
    \chi_{R_\bfv} = \chi_{R_{\varphi}}-1 = (1-D^2)P^{+}(D,g)\chi_{(0,0)}-1\,.
\end{equation}
    \item The spurion $\bft$ in the HEFT characterise the custodial symmetry breaking properties, which is the vev of degree of freedom and never participates into the Lorentz structure. Besides, the gauge structures with spurion self-contractions is redundant since such contractions gives nothing but an irrelevant constant. As discussed in Sec.~\ref{sec:spurion}, this redundancy can be eliminated by factorising out the Hilbert series of the spurion from the complete one, if the infinite series is truncated, such factorisation is just a subtraction, as shown in Eq.~\ref{eq:subs}.
    \item The physical Higgs field $h(x)$ is dimensionless since it always appears in the dimensionless polynomials $\mathcal{F}(h)$ in the effective operators. As a consequence, the number of the Higgs in an specific effective operator is arbitrary. Nevertheless, the number of independent operators does not increase endlessly when the number of Higgs becomes large. For the coverage of all the independent operators, we can truncate the infinite series at a relatively high power of $h(x)$, and the explicit value of such power is empirical and can be determined in practice. 
\end{itemize}
With the managements above, we can obtain the number of independent and complete effective operators of HEFT up to arbitrarily high dimension. The full results up to $d_\chi=10$ are listed in Table~\ref{tab:numbers}, besides, the result of the case where the spurion $\bft$ is of chiral dimension 1 is also presented in Table~\ref{tab:my_label}. The complete result of $d_\chi=3$ takes the form that
\begin{align}
    HS_{d_\chi=3} &= {n_f}^2h^3{\bfv}{L_R}{L_R^\dagger} + {n_f}^2h^3{\bfv}{L_L}{L_L^\dagger} + {n_f}^2h^3{\bfv}{Q_R}{Q_R^\dagger}+ {n_f}^2h^3{\bfv}
      {Q_L}{Q_L^\dagger} + 2{n_f}^2{\bft}h^3{\bfv}{L_R}{L_R^\dagger} \notag \\
      &+ 2{n_f}^2{\bft}h^3{\bfv}{L_L}{L_L^\dagger} + 2{n_f}^2{\bft}h^3{\bfv}
      {Q_R}{Q_R^\dagger} + 2{n_f}^2{\bft}h^3{\bfv}{Q_L}{Q_L^\dagger} + {n_f}^2{\bft}^2h^3{\bfv}{L_R}{L_R^\dagger} \notag \\
      &+ {n_f}^2{\bft}^2h^3{\bfv}{L_L}{L_L^\dagger} + {n_f}^2{\bft}^2h^3{\bfv}{Q_R}{Q_R^\dagger} + {n_f}^2{\bft}^2h^3{\bfv}{Q_L}{Q_L^\dagger}\,, \label{eq:hsd=3}
\end{align}
where $n_f$ is the flavor number, and the universal power of the Higgs $h(x)$ is taken to be 3 here. 
In particular, the terms with different numbers of $\bft$ are classified explicitly.
When the chiral dimension goes higher, the Hilbert series becomes complicated and it is tedious to list them all here, thus we collect the results up to $d_\chi=10$ in a txt document attached in the source: HEFTD10.txt. The detailed classifications of the operators from $d_\chi=3$ to $d_\chi=6$ can be found in Ref.~\cite{Sun:2022ssa,Sun:2022snw}.

\begin{table}
\centering
\resizebox{\textwidth}{!}{
\begin{tabular}{c|c|c|c}
\hline
$d_\chi$ & $\mathcal{N}_{\text{operator}}$ & $n_f=1$ & $n_f=3$ \\
\hline
3 & $16{n_f}^2$ & 16 & 144 \\
\hline
4 & $125 {n_f}^4-2 {n_f}^3+117 {n_f}^2+33$ & 273 & 11157 \\ 
\hline
5 & $\frac{4}{3} {n_f}^2 \left(575 {n_f}^2+3 {n_f}+340\right)$ & 1224 & 66288 \\
\hline
6 & \begin{tabular}{c}$\frac{1}{9} (17150 {n_f}^6-24 {n_f}^5+50015 {n_f}^4$\\$+258 {n_f}^3+21467 {n_f}^2+3726)$\end{tabular} & 10288 & 1861292 \\
\hline
7 & \begin{tabular}{c}$\frac{2}{9} {n_f}^2 (94325 {n_f}^4+372 {n_f}^3+151379 {n_f}^2$\\$+1104 {n_f}+46184)$\end{tabular} & 65192 & 18124552 \\
\hline
8 & \begin{tabular}{c} $\frac{1}{18} (453789 {n_f}^8-4116 {n_f}^7+3445400 {n_f}^6$\\$+12264 {n_f}^5+3565703 {n_f}^4$\\$+26124 {n_f}^3+891320 {n_f}^2+112986)$\end{tabular} & 472415 & 321147047 \\
\hline
9 & \begin{tabular}{c} $\frac{4}{45} {n_f}^2 (4602717 {n_f}^6-87269 {n_f}^5$\\$+16523580 {n_f}^4+50830 {n_f}^3$\\$+12294138 {n_f}^2+86344 {n_f}+2417430)$\end{tabular} & 3190024 & 3829824336 \\
\hline
10 & \begin{tabular}{c} $\frac{1}{225} (63392868 {n_f}^{10}-3810240 {n_f}^9$\\$+1050088850 {n_f}^8-22937520 {n_f}^7$\\$+2305118919 {n_f}^6+4351290 {n_f}^5+1315821375 {n_f}^4$\\$+8708370 {n_f}^3+221625088 {n_f}^2+21318300)$\end{tabular} & 22060788 & 54658127796 \\
\hline
\end{tabular}
}
\caption{The numbers of HEFT operators up to chiral dimension 10 of the case the spurion is of $d_\chi=0$.}
\label{tab:numbers}
\end{table}

\begin{table}[]
    \centering
    \resizebox{\textwidth}{!}{
    \begin{tabular}{c|c|c|c}
\hline
$d_\chi$ & $\mathcal{N}_{\text{operator}}$ & $n_f=1$ & $n_f=3$ \\
\hline
3 & $4{n_f}^2$ & 4 & 36 \\
\hline
4 & $\frac{1}{3}(125{n_f}^4+145{n_f}^2+42)$ & 104 & 3824 \\
\hline
5 & $\frac{525}{2}{n_f}^4-{n_f}^3+\frac{315}{2}{n_f}^2+9$ & 428 & 22662 \\
\hline
6 & \begin{tabular}{c}
     $\frac{1}{18}(8575{n_f}^6-6{n_f}^5+30373{n_f}^4+42{n_f}^3$\\
     $+13900{n_f}^2+1800)$
\end{tabular} & 3038 & 490998 \\
\hline
7 & \begin{tabular}{c}
     $\frac{1}{6}(32585{n_f}^6+58{n_f}^5+57949{n_f}^4+254{n_f}^3$ \\
     $+18000{n_f}^2 + 828)$
\end{tabular} & 18279 & 4772019 \\
\hline
8 & \begin{tabular}{c}
     $\frac{1}{90}(453789{n_f}^8 -2744{n_f}^7+4264760{n_f}^6+8500{n_f}^5$ \\
     $+4784006{n_f}^4+22414{n_f}^3+1219325{n_f}^2+105750)$
\end{tabular} & 120620 & 72017482 \\
\hline
9 & \begin{tabular}{c}
    $\frac{424977}{5}{n_f}^8-1078{n_f}^7+\frac{3153145}{9}{n_f}^6+608{n_f}^5$ \\
    $+\frac{12496772}{45}{n_f}^4 + 1368{n_f}^3+\frac{499285}{9}{n_f}^2+2061$ 
\end{tabular} & 771486 & 833882194 \\
\hline
10 & \begin{tabular}{c}
    $\frac{1}{450}(21130956{n_f}^{10}-952560{n_f}^9 + 431074500{n_f}^8-6962760{n_f}^7$ \\
    $+1055854043{n_f}^6+1152210{n_f}^5+635718350{n_f}^4+3081660{n_f}^3$ \\
    $107564951{n_f}^2+6726150)$ 
\end{tabular} & 5009750 & 10810252460 \\
\hline
    \end{tabular}
    }
    \caption{The numbers of HEFT operators up to chiral dimension 10 of the case the spurion is of $d_\chi=1$.}
    \label{tab:my_label}
\end{table}

\begin{figure}
\centering
  \includegraphics[scale=0.4]{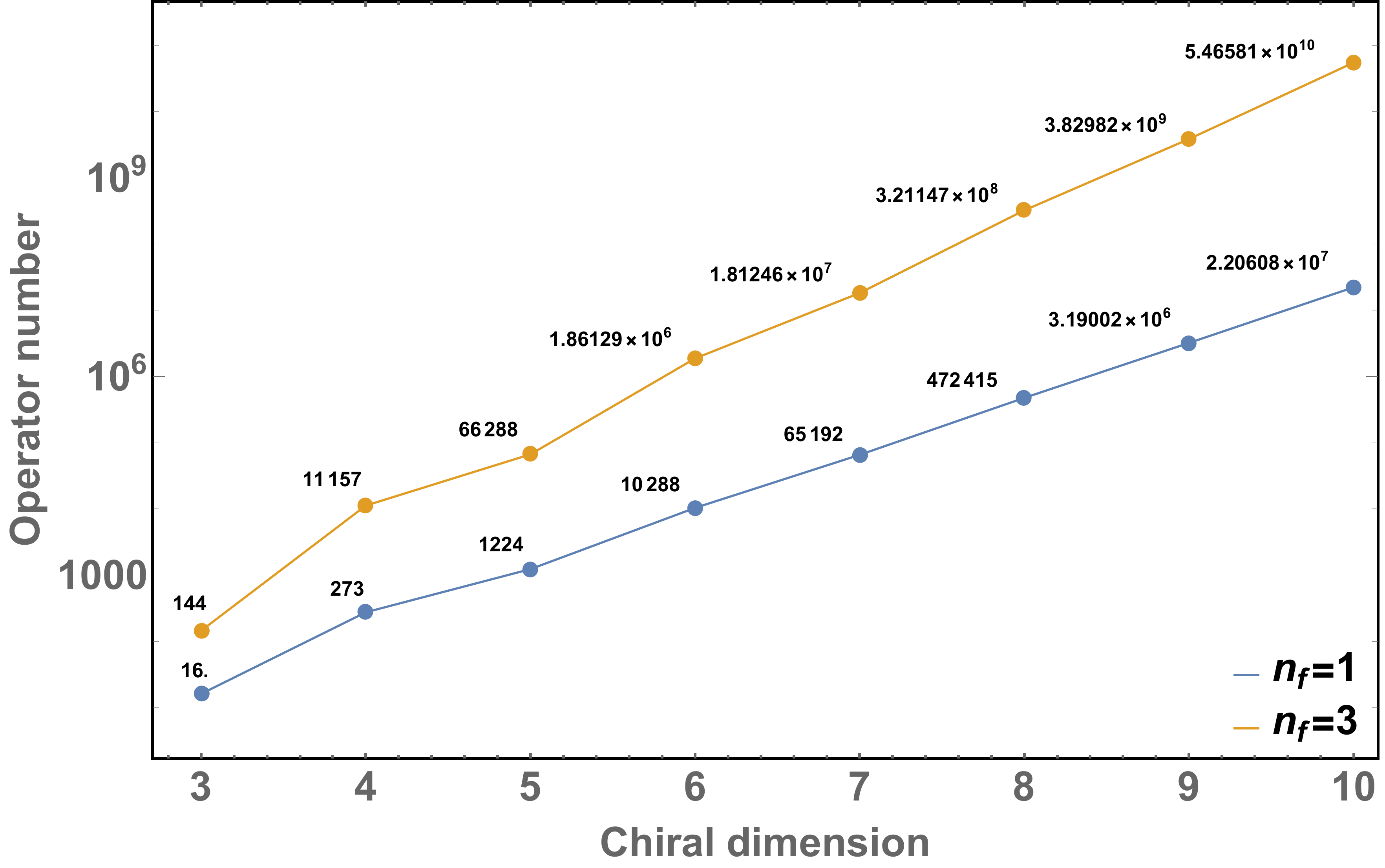}
  \caption{The growth of the number of the operators in the HEFT in the case of dimensionless spurion.}
  \label{graph:pig3}
\end{figure}

\begin{figure}
\centering
  \includegraphics[scale=0.4]{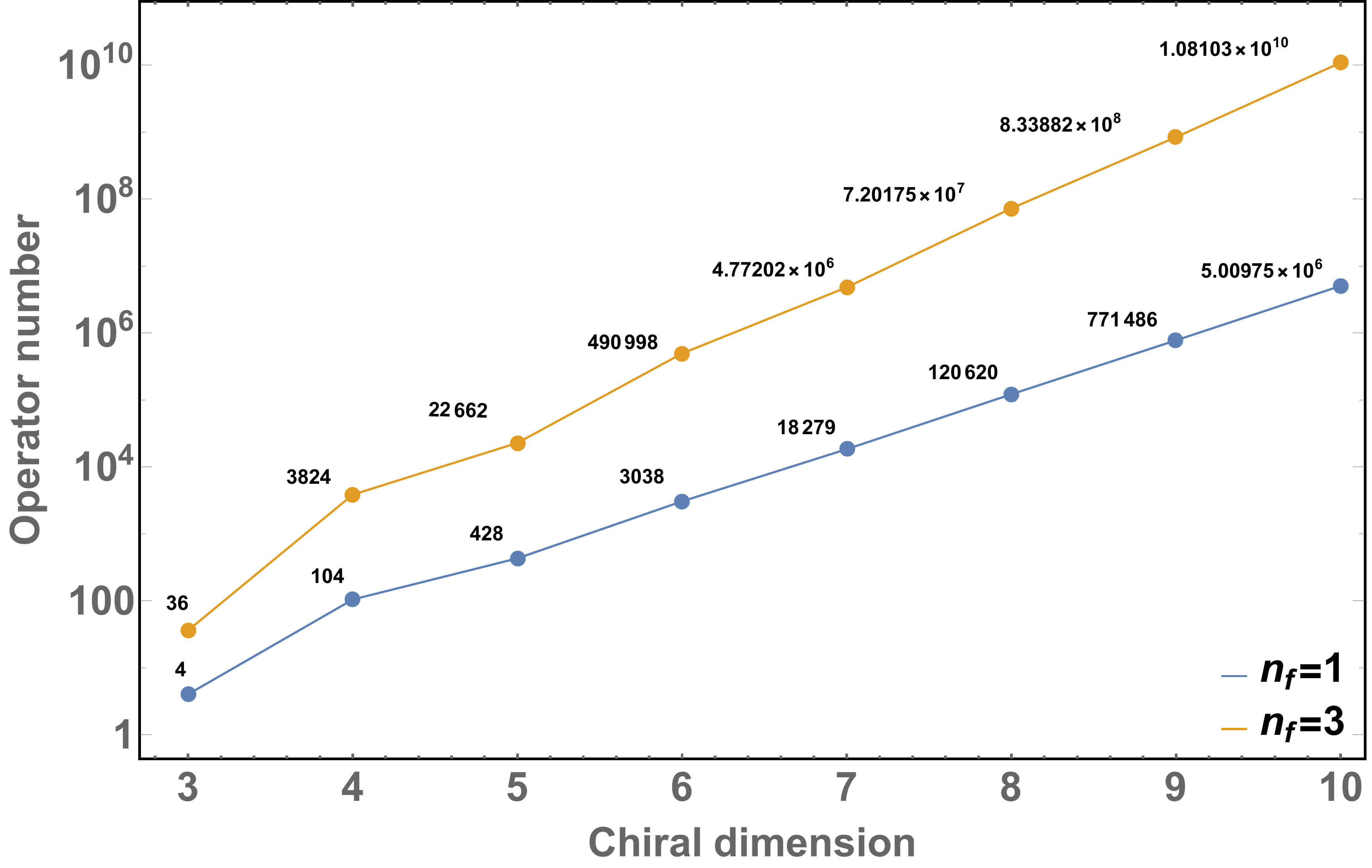}
  \caption{The growth of the number of the operators in the HEFT in the case of dimension-1 spurion.}
  \label{graph:pig4}
\end{figure}

The operator number of the HEFT increases rapidly, as shown in Fig.~\ref{graph:pig3} and Fig.~\ref{graph:pig4}, and the number of operators at chiral dimension 10 has been larger than the number of the SMEFT operators at the canonical dimension 15~\cite{Henning:2015alf}. 

\subsection{CP Property of the NLO Operators}

The effective operators can be further classified based on their CP properties, such that, to identify the CP violation effects. 
In the HEFT, the CP-even and CP-odd effective operators in the bosonic sector has been discussed in some literature such as Ref.~\cite{Buchalla:2013rka,Brivio:2016fzo}.
The Hilbert series method can also be utilized to analyse the CP nature of the effective operators. Although Ref.~\cite{Graf:2020yxt} discussed the CP properties of the bosonic fields, it is still non-trivial to extend the discussion on the operators involving in the fermionic fields. As shown in Sec.~\ref{sec:cp}, we have extended our discussion to the CP property of general fields and identify the numbers of the operators with specific CP nature.   


To obtain that, we need to combine different SPMs together to form the C, or P invariant space, correspondingly, the characters should be extended as well, as summarized in Table~\ref{CPtable2}. The characters of 4 different branches take different forms that
\begin{align}
    \chi_B^{C\pm P\pm} &= \chi^{C\pm P\pm}_{B_L}+\chi^{C\pm P\pm}_{B_R} \notag \\
    \chi_W^{C\pm P\pm} &= \chi^{C\pm P\pm}_{W_L}+\chi^{C\pm P\pm}_{W_R} \notag \\
    \chi_G^{C\pm P\pm} &= \chi^{C\pm P\pm}_{W_L}+\chi^{C\pm P\pm}_{W_R} \notag \\
    \chi_L^{C\pm P\pm} &= \chi^{C\pm P\pm}_{L_L}+\chi^{C\pm P\pm}_{L_R} + \chi^{C\pm P\pm}_{L_L^\dagger} + \chi^{C\pm P\pm}_{L_R^\dagger} \notag \\
    \chi_Q^{C\pm P\pm} &= \chi^{C\pm P\pm}_{Q_L}+\chi^{C\pm P\pm}_{Q_R} + \chi^{C\pm P\pm}_{Q_L^\dagger} + \chi^{C\pm P\pm}_{Q_R^\dagger} \,.
\end{align}
In the $C^+P^+$ branch, the characters above take the form in Table~\ref{tab:buildingblocks}, and the Hilbert series in this branch gives the total number of independent operators, while in other branches, the characters should be modified according to Table~\ref{CPtable2}. At the same time, the Haar measure also need modifications according to Table~\ref{CPtable1}.

For the NLO Lagrangian of the HEFT, which contains effective operators of $d_\chi=3,4$, the Hilbert series in the four branches are 
\begin{align}
    HS^{C+P+}_{d_\chi=3,4} &= 2 B^2+2 B L^2 {\bft}+2 B L^2+2 B Q^2 {\bft}+2 B Q^2+2 B {\bft} {\bfv}^2+2 B {\bft} W+D^4+D^3 {\bft} {\bfv}\notag \\
    &+2 D^2 L^2 {\bft}+2 D^2 L^2+2 D^2 Q^2 {\bft}+2 D^2 Q^2+2 D^2 {\bft}^2 {\bfv}^2+2 D^2 {\bfv}^2+4 D L^2 {\bft}^2 {\bfv}\notag \\
    &+8 D L^2{\bft} {\bfv}+4 D L^2 {\bfv}+4 D Q^2 {\bft}^2 {\bfv}+8 D Q^2 {\bft} {\bfv}+4 D Q^2 {\bfv}+D {\bft}^3 {\bfv}^3+D {\bft}^2 {\bfv}^3\notag \\
    &+2 D {\bft} {\bfv}^3+2 G^2+2 G Q^2 {\bft}+2 G Q^2+5 L^4 {\bft}^2+7 L^4 {\bft}+8 L^4+10 L^2 Q^2 {\bft}^2\notag \\
    &+30 L^2 Q^2 {\bft}+20L^2 Q^2+2 L^2 {\bft}^3 {\bfv}^2+6 L^2 {\bft}^2 {\bfv}^2+2 L^2 {\bft}^2 W+8 L^2 {\bft} {\bfv}^2+4 L^2 {\bft} W\notag \\
    &+4 L^2 {\bfv}^2+2 L^2 W+8 L Q^3 {\bft}+8 L Q^3+8 Q^4 {\bft}^2+12 Q^4 {\bft}+16 Q^4+2 Q^2 {\bft}^3 {\bfv}^2\notag \\
    &+6 Q^2 {\bft}^2{\bfv}^2+2 Q^2 {\bft}^2 W+8 Q^2 {\bft} {\bfv}^2+4 Q^2 {\bft} W+4 Q^2 {\bfv}^2+2 Q^2 W+{\bft}^4 {\bfv}^4\notag \\
    &+2 {\bft}^2 {\bfv}^4+2 {\bft}^2 {\bfv}^2 W+2 {\bft}^2 W^2+2 {\bft} {\bfv}^2 W+2 {\bfv}^4+2 {\bfv}^2 W+2 W^2 \notag \\
    &+2L^2\bfv + 4L^2\bfv\bft + 2L^2\bfv\bft^2 +2Q^2\bfv + 4Q^2\bfv\bft + 2Q^2\bfv\bft^2 \\
    HS^{C+P-}_{d_\chi=3,4} &= D^4-D^3 {\bft} {\bfv}+2 D^2 {\bft}^2 {\bfv}^2+2 D^2 {\bfv}^2-D {\bft}^3 {\bfv}^3-D {\bft}^2 {\bfv}^3-2 D {\bft} {\bfv}^3+L^4 {\bft}^2\notag \\
    &-L^4 {\bft}+2 L^4+2 Q^4 {\bft}^2-2 Q^4 {\bft}+4 Q^4+{\bft}^4 {\bfv}^4+2 {\bft}^2 {\bfv}^4+2 {\bfv}^4 
\end{align}
\begin{align}
    HS^{C-P-}_{d_\chi=3,4} &= 2 B^2+2 B L^2 {\bft}-2 B L^2+2 B Q^2 {\bft}-2 B Q^2+2 B {\bft} {\bfv}^2+2 B {\bft} W+D^4+D^3 {\bft} {\bfv}\notag \\
    &-2 D^2 L^2 {\bft}+2 D^2 L^2-2 D^2 Q^2 {\bft}+2 D^2 Q^2+2 D^2 {\bft}^2 {\bfv}^2+2 D^2 {\bfv}^2+D {\bft}^3 {\bfv}^3\notag \\
    &+D {\bft}^2 {\bfv}^3+2D {\bft} {\bfv}^3+2 G^2+2 G Q^2 {\bft}-2 G Q^2+3 L^4 {\bft}^2-3 L^4 {\bft}+6 L^4+6 L^2 Q^2 {\bft}^2\notag \\
    &-6 L^2 Q^2 {\bft}+12 L^2 Q^2-2 L^2 {\bft}^3 {\bfv}^2+2 L^2 {\bft}^2 {\bfv}^2+2 L^2 {\bft}^2 W-4 L^2 {\bft} {\bfv}^2\notag \\
    &+4 L^2 {\bfv}^2+2L^2 W+6 Q^4 {\bft}^2-6 Q^4 {\bft}+12 Q^4-2 Q^2 {\bft}^3 {\bfv}^2+2 Q^2 {\bft}^2 {\bfv}^2+2 Q^2 {\bft}^2 W\notag \\
    &-4 Q^2 {\bft} {\bfv}^2+4 Q^2 {\bfv}^2+2 Q^2 W+{\bft}^4 {\bfv}^4+2 {\bft}^2 {\bfv}^4+2 {\bft}^2 {\bfv}^2 W+2 {\bft}^2 W^2\notag \\
    &+2 {\bft} {\bfv}^2 W+2 {\bfv}^4+2{\bfv}^2 W+2 W^2 \\
    HS^{C-P+}_{d_\chi=3,4} &=  D^4-D^3 {\bft} {\bfv}+2 D^2 {\bft}^2 {\bfv}^2+2 D^2 {\bfv}^2-D {\bft}^3 {\bfv}^3-D {\bft}^2 {\bfv}^3-2 D {\bft} {\bfv}^3+3 L^4 {\bft}^2\notag \\
    &-3 L^4 {\bft}+4 L^4+4 L^2 Q^2 {\bft}^2-4 L^2 Q^2 {\bft}+8 L^2 Q^2+4 Q^4 {\bft}^2-4 Q^4 {\bft}+8 Q^4\notag \\
    &+{\bft}^4 {\bfv}^4+2{\bft}^2 {\bfv}^4+2 {\bfv}^4+2L^2\bfv+2L^2\bfv\bft^2+2Q^2\bfv+2Q^2\bfv\bft^2\,,
\end{align}
where the arbitrary $h$ factors has been dropped and the chiral components of the gauge bosons and fermions are collected
\bea
    \psi_L\,,\psi_R\,,\psi_L^\dagger\,,\psi_R^\dagger&\rightarrow & \psi\,,\quad \psi=L\,,Q \notag \\
    X_L\,,X_R&\rightarrow & X\,,\quad X=B\,,G\,,W\,.
\eea
For example, the type $L^2\bfv$ in $ HS^{C+P+}_{d_\chi=3,4}$ contains the types $h^3\bfv L_L L^\dagger_L\,,h^3\bfv L_R L^\dagger_R$ in the Hilbert series in Eq.~\ref{eq:hsd=3}.

Combining the four branches together, we can find the Hilbert series for the CP-even operators,
\begin{align}
     HS^{\text{CP-even}}_{d_\chi=3,4} &= B^2+B {\bft} {\bfv}^2+B {\bft} W+D^4+2 D^2 {\bft}^2 {\bfv}^2+2 D^2 {\bfv}^2+G^2 \notag \\
     &+{\bft}^4 {\bfv}^4+2 {\bft}^2 {\bfv}^4+{\bft}^2 {\bfv}^2 W+{\bft}^2 W^2+{\bft} {\bfv}^2 W+2 {\bfv}^4+{\bfv}^2 W+W^2 \notag \\
    &+B L^2 {\bft}+B Q^2 {\bft}+D^2 L^2+D^2 Q^2+D L^2 {\bft}^2 {\bfv}+2 D L^2 {\bft} {\bfv}+D L^2 {\bfv}+D Q^2 {\bft}^2 {\bfv}\notag \\
    &+2 D Q^2 {\bft} {\bfv}+D Q^2 {\bfv}+G Q^2 {\bft}+3 L^4{\bft}^2+5 L^4+5 L^2 Q^2 {\bft}^2+5 L^2 Q^2 {\bft}+10 L^2 Q^2\notag \\
    &+2 L^2 {\bft}^2 {\bfv}^2+L^2 {\bft}^2 W+L^2 {\bft} {\bfv}^2+L^2 {\bft} W+2 L^2 {\bfv}^2+L^2 W+2 L Q^3 {\bft}+2 L Q^3\notag \\
    &+5 Q^4 {\bft}^2+10 Q^4+2 Q^2 {\bft}^2 {\bfv}^2+Q^2{\bft}^2 W+Q^2 {\bft} {\bfv}^2+Q^2 {\bft} W+2 Q^2 {\bfv}^2\notag \\
    &+Q^2 W+ L^2 \bft^2 {\bfv}+L^2 \bft {\bfv}+L^2 {\bfv}+Q^2 \bft^2 {\bfv}+Q^2 \bft {\bfv}+Q^2 {\bfv}
\end{align}
where the first two lines are of bosonic sector. Comparing with the CP-even bosonic operators presented in the previous literature, there are some disagreements: there is another one type $D\bft^2\bfv^3$ in Ref.~\cite{Buchalla:2013rka} excluded in the result here, while in Ref.~\cite{Brivio:2016fzo}, there are also extra 4 types
\begin{equation}
    D\bft^2\bfv^3\,, B\bft\bfv D\,,W\bfv D\,,   W\bft\bfv D
\end{equation}
which are not counted by the Hilbert series here. 
\section{Conclusion}
\label{sec:conc}

In this work we utilize the Hilbert series method to count the numbers of the complete and independent operators in the Higgs EFT. The redundancies from equation of motion and integration-by-part in effective operators can be systematically removed by modifying the 
characters of the single particle module and extracting out the primary operators from the operator set. We present the following new treatments appeared in the HEFT operator counting:
\bit

\item The operators involving in spurions has additional redundancy, which can be eliminated by the factorisation property of the Hilbert series. The physical Hilbert series is the quotient of the complete Hilbert series divided by the one composed solely by the spurions. 
\item The CP property is incorporated by extending the symmetry group to its outer automurphism. We generalise the CP properties of the bosonic field discussed in Ref.~\cite{Graf:2020yxt} to include fermion fields, which is a non-trivial extension.
\eit

With the above new treatments, we present the Hilbert series of the HEFT up to chiral dimension 10. An efficient FORM code is also provided, which can be utilized to obtain the counting results to higher and higher orders. Furthermore, we also show four different branches of the Hilbert series at the NLO Lagrangian, including the CP-even, and CP-odd operators. Finally we would like to mention that the procedure in this work is universal and thus it can be applied to other EFTs with the Goldstone bosons or spurions included.

%
%
%


\section*{Note added}

While this work was being completed, a related work~\cite{Graf:2022rco} appears on the arXiv. Ref.~\cite{Graf:2022rco} counts the Higgs EFT operators at the NLO, while we present the counting results up to chiral dimension 10. In particular, the NLO operator countings in this work are consistent with the ones of the so-called $\nu$HEFT in~\cite{Graf:2022rco}. Furthermore, we also discuss the CP properties of the HEFT operators and list the numbers of the CP-even and CP-odd operators at the NLO.

\section*{Acknowledgments}

The authors are supported by the National Science Foundation of China under Grants No. 12022514, No. 11875003 and No. 12047503, and National Key Research and Development Program of China Grant No. 2020YFC2201501, No. 2021YFA0718304, and CAS Project for Young Scientists in Basic Research YSBR-006, the Key Research Program of the CAS Grant No. XDPB15.

\bibliography{ref}
\end{document}